\documentclass[sigplan,10pt,screen]{acmart}

\usepackage[T1]{fontenc}
 \usepackage{textcomp, libertine}
\setcopyright{none}
\acmPrice{}
\acmDOI{}
\acmYear{2021}
\acmISBN{}
\acmConference[pe-graph2axiom]{pe-graph2axiom}{June, 2021}{ArXiV}
\usepackage{booktabs} % For formal tables

\usepackage{courier}            % standard fixed width font
\usepackage[scaled]{helvet} % see www.ctan.org/get/macros/latex/required/psnfss/psnfss2e.pdf
\usepackage{url}                  % format URLs
\usepackage{paralist}
\usepackage{enumitem}      % adjust spacing in enums
\usepackage{breakurl}
\usepackage{amsfonts}
\usepackage{subfig}
\usepackage{subfloat}
\usepackage{stfloats}
\usepackage{clrscode}
\usepackage{xcolor}
\usepackage{listings}
\usepackage{tikz}

\usepackage[linesnumbered,ruled]{algorithm2e}

\usepackage{graphicx} 	% For figures
\usepackage{mathtools}	% For math function

\usepackage{multirow}	% For table

\usepackage{amsmath}
\usepackage{amsthm}

\usepackage{bm}
\usepackage{xspace}
\newcommand{\ignore}[1]{}

\definecolor{mygreen}{rgb}{0,0.6,0}
\definecolor{mygray}{rgb}{0.5,0.5,0.5}
\definecolor{mymauve}{rgb}{0.58,0,0.82}

\usepackage{caption}
\usepackage{rotating}
\usepackage{cancel}

\usepackage{balance}

\usepackage{ragged2e}
\usepackage{array}
\newcolumntype{R}[1]{>{\RaggedLeft\arraybackslash}p{#1}}
\newcolumntype{P}[1]{>{\centering\arraybackslash}p{#1}}

\usepackage[section]{placeins}

\begin{document}

%% Title information
%% \title[PEGraph2Seq: Robust Program Equivalence Checking using Machine Learning]{PEGraph2Seq: Robust Program Equivalence Checking\\ using Machine Learning}

%%% LNP: tentative?
\title[Proving Equivalence Between Complex Expressions]{Proving Equivalence Between Complex Expressions\\ Using Graph-to-Sequence Neural Models}

%% [Short Title] is optional;
                                        %% when present, will be used in
                                        %% header instead of Full Title.
%\titlenote{with title note}             %% \titlenote is optional;
                                        %% can be repeated if necessary;
                                        %% contents suppressed with 'anonymous'
%\subtitle{Subtitle}                     %% \subtitle is optional
%\subtitle{\vspace{-.4cm}{\bf Under review (anonymous) at ACM PLDI'20 -- DO NOT DISTRIBUTE}\vspace{-.1cm}}
%\subtitlenote{with subtitle note}       %% \subtitlenote is optional;
                                        %% can be repeated if necessary;
                                        %% contents suppressed with 'anonymous'

%% Author information
%% Contents and number of authors suppressed with 'anonymous'.
%% Each author should be introduced by \author, followed by
%% \authornote (optional), \orcid (optional), \affiliation, and
%% \email.
%% An author may have multiple affiliations and/or emails; repeat the
%% appropriate command.
%% Many elements are not rendered, but should be provided for metadata
%% extraction tools.

%% Author with single affiliation.
\author{Steve Kommrusch, Th{\'e}o Barollet and Louis-No{\"e}l Pouchet}
\begin{abstract}
%% \begin{abstract}
%% \input{abstract}
%% \end{abstract}

We target the problem of provably computing the equivalence between two complex expression trees. To this end, we formalize the problem of equivalence between two such programs as finding a set of semantics-preserving rewrite rules from one into the other, such that after the rewrite the two programs are structurally identical, and therefore trivially equivalent. We then develop a graph-to-sequence neural network system for program equivalence, trained to produce such rewrite sequences from a carefully crafted automatic example generation algorithm. We extensively evaluate our system on a rich multi-type linear algebra expression language, using arbitrary combinations of 100+ graph-rewriting axioms of equivalence. Our machine learning system guarantees correctness for all true negatives, and ensures 0 false positive by design. It outputs via inference a valid proof of equivalence for 93\% of the 10,000 equivalent expression pairs isolated for testing, using up to 50-term expressions. In all cases, the validity of the sequence produced and therefore the provable assertion of program equivalence is always computable, in negligible time.

\end{abstract}

%% 2012 ACM Computing Classification System (CSS) concepts
%% Generate at 'http://dl.acm.org/ccs/ccs.cfm'.

%% End of generated code

%% Keywords
%% comma separated list
%\keywords{Program equivalence, deep learning, graph neural network, graph-to-sequence systems}  %% \keywords are mandatory in final camera-ready submission

%% \maketitle
%% Note: \maketitle command must come after title commands, author
%% commands, abstract environment, Computing Classification System
%% environment and commands, and keywords command.
\maketitle
%% \vspace{.8cm}

\section{Introduction}
\label{sec:introduction}
%% \section{Introduction}
%% \label{sec:introduction}
%% \input{introduction}

Deep neural network systems have excelled at a variety of classification and reinforcement learning tasks \cite{Goodfellow16}. However, their stochastic nature tends to hinder their deployment for automated program analysis: ensuring the correctness of the solution produced is often required, e.g., when determining the semantics equivalence between two programs (or symbolic expressions).

In this work we target the problem of automatically computing whether two input symbolic expressions are semantically equivalent \cite{kaplan1969regular}, under a well-defined axiomatic system for equivalence using semantics-preserving rewrite rules \cite{dershowitz1985computing}. Program equivalence is summarized as determining whether two programs would always produce the same outputs for all possible inputs, and is a central problem in computing \cite{kaplan1969regular,godlin2008inference,verdoolaege2009equivalence}. The problem ranges from undecidable, e.g. \cite{goldblatt2012well}, to trivial in cases of testing the equivalence of a program with itself. Our work directly studies the subset of programs represented by symbolic linear algebra expressions which include scalar, vector, and matrix types for both constants and variables, and 16 different operators with 147 distinct axioms of equivalence. For example, the expression using matrices, scalars, and a vector: $(A+B)I((a+(b-b))/a)\vec v - A\vec v$ can be proven equivalent to $B\vec v$ by applying 10 axioms in sequence; our work generates the proof steps between these expressions.

While prior work has shown promises for deep networks to compute some forms of program equivalence \cite{Xu17,Alon19}, the system typically outputs only a probability of equivalence, without any reasoning or insight that can be verified easily: false positive can be produced. Programs can be represented as a tree (or graph) of symbols, and deep networks for symbolic reasoning have been studied, e.g. to compute the derivative of a symbolic expression \cite{Lample20}. In this work, we take a significantly different approach to the problem of symbolic program reasoning with deep networks: we make the system produce the sequence of steps that lead to rewriting one program into another, that is the \emph{reasoning} for (or proof of) equivalence between the two programs, instead of producing directly the result of this reasoning (e.g., a probability of equivalence, without explanation about the reasoning). In a nutshell, we approach expression equivalence as a theorem proving problem, in which all the axioms as well as tactics to compute a proof are all learned by example in a deep learning system, without any human insight.

%This proof, made of the sequence of axioms and their point of application on the program, can be automatically and trivially verified for correctness, by simply checking whether the axioms can be legally applied as suggested

%, targeting explicitly the automated and systematic verification of the network output, to 

We propose a method for generating training samples using probabilistic applications of production rules within a formal grammar, and then develop a graph-to-sequence \cite{Li16,Beck18} neural network system for program equivalence, trained to learn and combine rewrite rules to rewrite one program into another.
It can \emph{deterministically} prove equivalence, entirely avoids false
positives, and quickly invalidates incorrect answers produced by the network
(no deterministic answer is provided in this case, only a probability of non-equivalence). In a
nutshell, we develop the first graph-to-sequence neural network system to
accelerate the search in the space of possible combinations of
transformation rules (i.e., axioms of equivalence in the input
language) to make two graphs representing symbolic expressions structurally identical without violating their original semantics. 
We propose a machine learning system for program equivalence which ensures correctness for all non-equivalent programs input (specificity = 100\%) , and a deterministically checkable output for equivalent programs (no false positives). We make the following contributions:
\begin{enumerate}%[noitemsep,topsep=0pt,wide=0pt]
\item We design, implement and evaluate two competing approaches using graph-to-sequence neural network systems to generate proofs of equivalence. We provide the first implementation of such graph-to-sequence systems in the popular OpenNMT-py framework \cite{opennmt}.

\item We present a complete implementation of our system operating on a rich language for multi-type linear algebra expressions. Our system provides a correct rewrite rule sequence between two equivalent programs for 93\% of the 10,000 test cases. The correctness of the rewrite rule is deterministically checkable in all cases in negligible time.
  
\end{enumerate}

The rest of the paper is organized as follows. Sec.~\ref{sec:motivation} outlines the program equivalence problem we address, and motivates our proposed approach. Sec.~\ref{sec:proglangdefs} formalizes the equivalence problem addressed. Automatic sample generation is discussed in Sec.~\ref{sec:samplegen} before Sec.~\ref{sec:progequivdnn} which introduces our DNN system, its overall design principles and key components. A complete experimental evaluation of our system is detailed in Sec.~\ref{sec:expresults}. We present related work in Sec.~\ref{sec:related} before concluding.

\section{Motivation and Overview}
\label{sec:motivation}

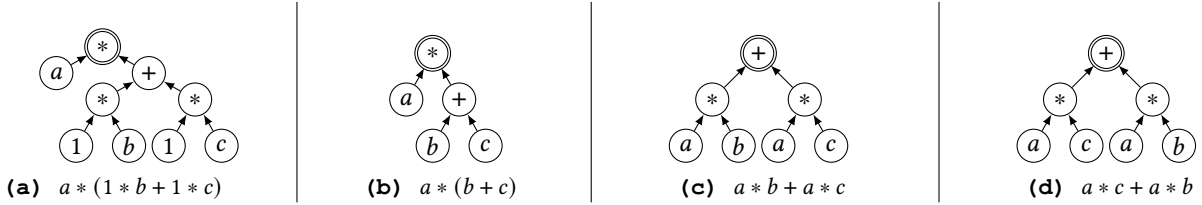
\begin{figure*}
    \centering

    %% \begin{subfigures}
    \centering
      \subfloat[][$a*(1* b+1* c)$]{
      \begin{minipage}[b]{0.25\textwidth}
        \label{fig:treeexamples:1}
        \centering
        \begin{tikzpicture} [level distance=1em, inner sep=1pt, minimum size=1.25em, edge from parent/.style={draw,latex-}]
            \node [circle, double, draw] {$*$}
            child [sibling distance=3.5em] {node [circle, draw] {$a$}
            child{edge from parent[draw=none] node [opacity=0] {}
            child  [sibling distance=2em]{edge from parent[draw=none] node [opacity=0] {}}
            child  [sibling distance=2em]{edge from parent[draw=none] node [opacity=0] {}}
            }
            child{edge from parent[draw=none] node [opacity=0] {}}
            }
            child [sibling distance=3.5em] {node [circle, draw] {$+$}
            child [sibling distance=3.5em, level distance=1em] {node [circle, draw] {$*$}
            child [sibling distance=2em, level distance=1.75em] {node [circle, draw] {$1$}}
            child [sibling distance=2em, level distance=1.75em] {node [circle, draw] {$b$}}
            }
            child [sibling distance=3.5em, level distance=1em] {node [circle, draw] {$*$}
            child [sibling distance=2em, level distance=1.75em] {node [circle, draw] {$1$}}
            child [sibling distance=2em, level distance=1.75em] {node [circle, draw] {$c$}}
            }
            };
        \end{tikzpicture}
        %\caption{$a*(1* b+1* c)$}        
        \end{minipage}
      %% \end{subfigures}
      }
      \unskip\ \hfill \vrule \hfill
      \subfloat[][$a*(b+c)$]{
    %% \begin{subfigures}
      \begin{minipage}[b]{0.20\textwidth}
        \centering
        \begin{tikzpicture} [level distance=1.75em, inner sep=1pt, minimum size=1.25em, sibling distance=2em, edge from parent/.style={draw,latex-}]
            \node [circle, double, draw, align=center] {$*$}
                child {node [circle, draw] {$a$}
                    child{edge from parent[draw=none] node [opacity=0] {}}
                    child{edge from parent[draw=none] node [opacity=0] {}}
                }
                child {node [circle, draw] {$+$}
                    child{node [circle, draw] {$b$}}
                    child {node [circle, draw] {$c$}}
                };
        \end{tikzpicture}
        %% \caption{$a*(b+c)$}
        \label{fig:treeexamples:2}
        \end{minipage}
      %% \end{subfigures}
      }
    \unskip\ \hfill \vrule \hfill
    %% \begin{subfigures}
    \subfloat[][$a * b + a * c$]{
      \begin{minipage}[b]{0.24\textwidth}
        \centering
        \begin{tikzpicture} [level distance=1.75em, inner sep=1pt, minimum size=1.25em, sibling distance=3.5em, edge from parent/.style={draw,latex-}]
            \node [circle, double, draw] {$+$}
                child {node [circle, draw] {$*$}
                    child [sibling distance=2em] {node [circle, draw] {$a$}}
                    child [sibling distance=2em] {node [circle, draw] {$b$}}
                }
                child {node [circle, draw] {$*$}
                    child [sibling distance=2em] {node [circle, draw] {$a$}}
                    child [sibling distance=2em] {node [circle, draw] {$c$}}
                };
        \end{tikzpicture}
        %% \caption{$a * b + a * c$}
        \label{fig:treeexamples:3}
        \end{minipage}
      %% \end{subfigures}
      }
    \unskip\ \hfill \vrule \hfill
    \subfloat[][$a * c + a * b$]{
    %% \begin{subfigures}
      \begin{minipage}[b]{0.24\textwidth}
        \centering
        \begin{tikzpicture} [level distance=1.75em, inner sep=1pt, minimum size=1.25em, sibling distance=3.5em, edge from parent/.style={draw,latex-}]
            \node [circle, double, draw] {$+$}
                child {node [circle, draw] {$*$}
                    child [sibling distance=2em] {node [circle, draw] {$a$}}
                    child [sibling distance=2em] {node [circle, draw] {$c$}}
                }
                child {node [circle, draw] {$*$}
                    child [sibling distance=2em] {node [circle, draw] {$a$}}
                    child [sibling distance=2em] {node [circle, draw] {$b$}}
                };
        \end{tikzpicture}
        %% \caption{$a * c + a * b$}
        \label{fig:treeexamples:4}
      \end{minipage}
      }
    \caption{Examples of Computations}
    \label{fig:treeexamples}
    %% \end{subfigures}

\end{figure*}

%%%%%%%%%%%%%%%%%%
%%%%%%%%%%%%%%%%%%

\paragraph*{Rewrite rules as axioms of equivalence}
%\FIXME{Should we change this whole section to only discuss figure b through d (delete the (a) diagram to help compress? And then cut down some of the text.}
In this work we represent programs with symbolic expressions made of variables (e.g., $a$, $b$, $c$), operators (e.g., \texttt{+}, \texttt{*}) and neutral/absorbing elements (e.g., $1$). We consider a rich linear algebra expression language, supporting three variable types (scalars as shown in P1-P4, vectors, and matrices) and 5 different variables per type; 16 operators including operators mixing different variable types such as vector-matrix product. We represent these programs as dataflow graphs \cite{buck1993scheduling} with a single root node, that is to compute a single value.

P1 is equivalent to P2 if we consider the axiom $A1: 1_{\mathbb{N}} * x = x,~\forall x \in \mathbb{N}$. This axiom is also a clear rewrite rule: the LHS expression $1_{\mathbb{N}} * x$ (with $x\in\mathbb{N}$) can be matched and replaced by the RHS expression $x$ anywhere in the program without altering its semantics. An axiom, or equivalently here a graph rewrite rule, may be applied repeatedly to different subtrees. When applying $A1$ on a specific location, the node $b$ of $P1$, we obtain an equivalent and yet syntactically different program, we note $P1 \equiv A1(b,P1)$. These equivalences can be composed, incrementally, to form a complex transformation: we have $P1 \equiv A1(c,A1(b,P1))$. The result of these semantics-preserving transformations can be computed in sequence: first implement $A1(b,P1)$ to obtain a new program $P'$, then $A1(c,P')$ to obtain $P''$. To prove $P1 \equiv P2$, we simply check $P''$ is structurally identical to $P2$, a linear time process.

To assess the validity of a transformation sequence $S$ where $P2 = S(P1)$, one simply needs to check for $S$, in sequence, that each axiom is applicable at that program point, apply it to obtain a new temporary program, and repeat the process for each axiom in the complete sequence. If the sequence is verified to be valid, and $S(P1)$ is structurally equivalent to $P2$, then we have proved $P1 \equiv P2$, and $S$ forms the complete proof of equivalence between the two programs. Using  $A2:x * (y+z) = x*y+x*z,~\forall x,y,z \in \mathbb{N}$ and $A3:x+y = y+x,~\forall x,y \in \mathbb{N}$, we have 
$P1 \equiv P4 \equiv A3(+,A2(*,A1(c,A1(b,P1))))$, a verifiable proof of equivalence under our axioms between the programs $a(1b+1c)$ and $ac+ab$, which involved structural changes including node deletion, creation and edge modification.
Note the bidirectional nature of the process: one can rewrite from $a(1b+1c)$ to $ac+ab$, or the converse using the same (but reverted) sequence. Note also the non-unicity of a sequence: by possibly many ways a program can be rewritten into another one, for example the sequence $P4 \equiv A3(+,A1(c,A1(b,A2(*,P1))$ also correctly rewrites $P1$ into $P4$. Conversely, a sequence may not exist: for example no sequence of the 3 above axioms allow to rewrite $a+b$ into $a*b$. We call these non-equivalent in our system, that is precisely if there is no sequence of axioms that can be applied to rewrite one program into the other. 
Our approach aims to compute some $S$ for a pair of programs $P1,P2$, so that $S$ is verified correct when $P1 \equiv P2$. Consequently, if $P1 \not\equiv P2$, no sequence $S$ produced can be verified correct: true negatives are trivially detected. 
%The challenge we address in this work is: given two programs $P1,P2$, how to successfully generate $S$ if $P1 \equiv S(P2)$? 
%This approach e
%but ensuring we produce $\bot$ if $P1 \not\equiv P2$? 

\paragraph*{Pathfinding program equivalence proofs}
Intuitively, we can view the solution space as a graph, where every possible syntactically different program in the language is represented by its own vertex $v_i$. And
$\exists ~e^{({A_k},x)}: v_i \rightarrow v_j$ iff $\exists A_k$ an axiom and $x$ a node in $v_i$ such that $v_j = A_k(x,v_i)$.
%Two vertices are connected by an edge iff one is the result of the correct application of a single axiom on the other, labeled by that axiom and program point at which it is applied.
%
%
%There is a directed edge $e_{A_k}^x$ between two vertices $P_i,P_j$ iff $P_j$ is the program resulting from the valid application of axiom $A_k$ on node $x$ of program $P_i$. 
%
Any two programs connected by a path in this graph are therefore semantically equivalent. Building $S$ for $P1 \equiv S(P2)$ amounts to exposing one path between $P1$ and $P2$ in this graph when it exists, the path forming the proof of equivalence. We build a deep learning graph-to-sequence system to learn a stochastic approximation of an iterative algorithm to construct such feasible path when possible, trained only by randomly sampling pairs of programs and one carefully labeled path between them. This avoids the need to craft smart exploration heuristics to make this path-finding problem practical.

\begin{figure*}[h!tb]
\vspace{-.4cm}
\centering\includegraphics[width=14cm]{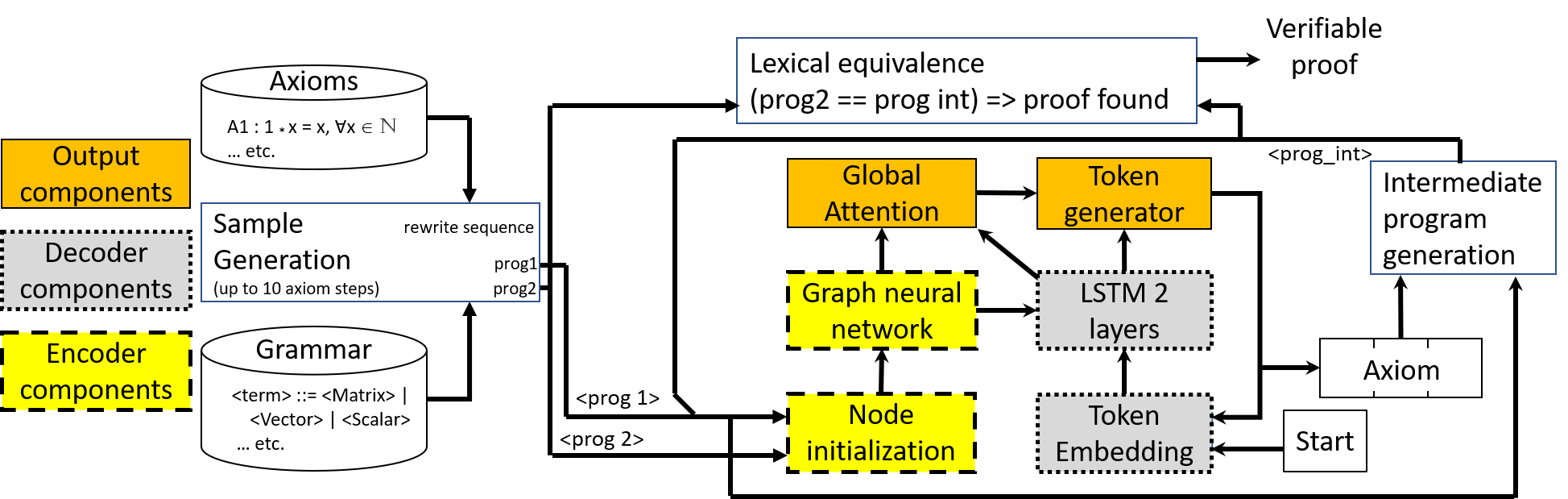}
\caption{\texttt{pe-graph2axiom} System Overview}
\vspace{-.4cm}
\label{fig:Full}
\end{figure*}

\paragraph*{Graph-to-sequence network for pathfinding}
This is instead what we let the neural network learn automatically; and specifically why we implemented graph neural networks to solve this problem \cite{Scarselli09,Xu17}. We rely on the network to suggest a transformation path by inference, and then verify its validity in linear time.
To implement our approach, we enumerate randomly valid sentences in a language, and a set of axioms of equivalence expressible as semantics-preserving rewrite rules from one to the other. 
The system in Fig.~\ref{fig:Full} takes as input two programs represented as symbolic trees, and produces a sequence of axioms along with their position of application (or node) that can be used to rewrite sequentially one input program into the other input program. 
%As each axiom is produced, it is checked to insure it is a legal application within the grammar and if the transformed program matches the target then a correct proof of equivalence has been found.
To train the system, we generate pairs of equivalent programs by iterating the axioms with random probability on one program, thereby generating both a path to equivalence and the target program. Random programs are generated so as to respect the grammar defined. The training set is then appropriately selected from these random samples, as detailed in Sec.~\ref{sec:expresults}.
%The system in Fig.~\ref{fig:Full}  is composed of the following blocks. 
\emph{Node initialization} initializes the graph neural network, converting the input programs text (e.g., $(a+(b+c))$ into nodes and edges in the \emph{Graph Neural Network} \cite{Scarselli09,Xu17}.
%refers to a neural network that has weights which allow it to learn interrelations between network nodes based on edge connections.
%We use \emph{Global attention} \cite{luong15} to connect tokens produced with nodes in the graph.
%allows the decoder to pay attention to certain nodes in the graph as it creates each token in the output sequence.
%
%A key of our approach is to introduce graph-to-sequence neural networks to quickly compute one or more possible rewrite sequences. 
%
The details of the network are covered in Sec.~\ref{sec:progequivdnn}. In a nutshell, the key principle is to combine a memory-based neural network approach, e.g., using Long-Short Term Memory (LSTM) \cite{Hochreiter97} neurons and a graph neural network design (which uses Gated Recurrent Units (GRUs) internally) \cite{Beck18} that matches our program graph representation.  \emph{Token embedding} is a neural network layer in which tokens are assigned a learnable multidimensional embedding vector \cite{Mikolov13}.
Each layer in \emph{LSTM 2 layers} has 256 neurons, which support sequence generation.
\emph{Token generator} is the final output portion of the network. It learns to output the tokens based on the current LSTM hidden states and the \emph{Global Attention} from the graph neural network. As each token is output, it feeds back into the LSTM layer through the embedding layer to affect its next state. We use a sequence 
generation principle, using a global attention mechanism \cite{luong15} to allow observation of program graph node information while generating the axiom and location on which it is applied. As developed below, we specifically study the robustness of our approach to generate proofs of increasingly complex length, contrasting models to output the entire path at once with \texttt{pe-graph2axiom} which incrementally builds the sequence one step at a time, as shown in Sec.~\ref{sec:expresults}.

\section{Framework for Program Equivalence}
\label{sec:proglangdefs}
%% \section{Programming Language}
%% \label{sec:proglangdefs}
%% \input{proglangdefs}

We now present the formalism we use in this work to represent symbolic expressions and their equivalences. We carefully co-designed this problem representation and the (graph) neural network approach to make the best use of machine learning via deep networks, as discussed in Sec.~\ref{sec:progequivdnn}. 

%% While our approach could be deployed on a larger class of problems, we stick to defining what we belive 

\subsection{Input Representation}

A key design aspect is to match the capability of the neural network to model the input as a walkable graph with the actual input program representation to be handled. We therefore model ``programs'' in a dataflow-like representation (i.e., a directed graph), using a single root/output node. Symbolic expressions computing a single result typically fit this representation. The following definitions are applicable to programs represented as dataflow graphs, albeit we specialize them to symbolic expressions.

%In particular, we do not restrict to tree-like structures nor acyclic graphs, as briefly discussed in Sec.~\ref{sec:ruminations-on-extensions}. 
%An input program graph is made of nodes and edges defined as follows.

\begin{definition}[Expression graph node]
  \label{def:proggraphnode}
A node $n \in N$ in the expression graph models n-ary operations and input operands. A node produces a value which can be consumed by any of its immediate successors in the graph. When a node has no predecessor, it models an input value. The output value for the computation is produced by the unique root node $n_{root}$ of the graph, the only node without successor.
\end{definition}

\begin{definition}[Expression graph directed edge]
  \label{def:proggraphedge}
  A directed edge $e_{n_1,n_2} : n_1 \rightarrow n_2$ with $n_1, n_2 \in N$ in the expression graph connects the producer of a value ($n_1$) to a node consuming this value in the computation. 
\end{definition}

\begin{definition}[Expression graph]
  \label{def:proggraph}
A expression graph $G$ is a directed dataflow graph modeling the computation, made of nodes $n_i \in N$ and edges $e_{n_i,n_j} \in E$ as defined in Def.~\ref{def:proggraphnode} and Def.~\ref{def:proggraphedge}. That is,  $G = \langle n_{root}, N, E \rangle$. There is no dandling edge nor unconnected node in $G$. 
\end{definition}

\paragraph*{Language of linear algebra expressions} We developed a complex-enough language to evaluate carefully our work, that captures rich linear algebra expressions. Specifically, we support 3 types of data/variables in the expression: scalars, vectors and matrices. We use the standard notation $a,\vec a,A$ for scalars, vectors and matrices.
%% The objective is to make the program complex enough to show reasoning based on operand/operation type, and avoid having a single operation type.
We evaluate using different variable names for each of the 3 types above, along with their identity and absorbing elements.
%% In other words, in our language there are 50 possible input values/variables that can be used in the program, from 3 different data types.
%% \FIXME{LNP: Why not 51? i.e., why not having 1 for vectors too?}

We also model a rich set of operators, mixing different unary and binary operations for each type. Specifically, we support $*_s,+_s,-_s,/_s$ between scalar operands, and $+_v,-_v,*_v$ between vectors and $+_m,-_m,*_m$ for matrices. For $-,/$ we also support their unary version for all types, e.g. $^{-1_{s}}$ for unary scalar inversion and $-_{um}$ for unary matrix negation. For example $a^{-1_s}$ computes to $1/a$.
We also support multi-type operations, such as vector and matrix scaling by a scalar $*_{sv}, *_{sm}$. We support two specific unary matrix operations, transpose $^{t_m}$ and matrix inversion as $^{-1_m}$. Note every operator has a unique name in our language, driven by the type of its operand. This will facilitate the learning of the expression embedding, avoiding the need to learn type propagation.

\paragraph*{Examples} Expressions of the form $A (B C^t D) E^{-1}$, $\vec a + b\vec c^{-1}-0\vec e$, $(a+b)+(c(d/e))$, $(aA+bB)C^t$ etc. can be parsed trivially to our representation, one simply needs to be able to provide a unique name for each operand and operator type (possibly via some analysis, or simple language design principles), that is avoiding to overload the semantics of operators and operands. Note the semantics is never explicitly provided to our DNN approach, it is learned by examples. There will be no example of the form e.g. $a+A$, an invalid expression in our language.

We believe a sensible approach is to develop a clean, regular grammar for the language to be handled, as implicitly these are concepts the DNN will need to learn. We did so, using a classical LL(1) grammar description of our linear algebra language. This is not a requirement of our approach, as one can arrive to the desired input expression graph by any means necessary, but we believe making the reasoning on the language structure ``easy'' is an important design aspect.

\subsection{Axioms of Equivalence}

A central aspect of our approach is to view the problem of expression
equivalence as finding a sequence of locally-correct rewrite rules
that each preserve the semantics, \emph{thereby making incremental reasoning possible}. We explicitly do not
consider non-semantics-preserving axioms. A rich structure of alternate but
equivalent ways to rewrite one expression to another makes the problem
easier to sample and more amenable to machine learning. Semantics-preserving axioms enable incremental per-axiom reasoning, and enforce semantics preservation without overly complicated semantics analysis; while still manipulating a very
rich space of transformations. To illustrate this we specifically
design axioms that perform complex graph modifications, such as node
deletion or creation, subtree manipulation, multi-node graph changes,
etc. %We now define the rewrite rule system we developed. 

%% \FIXME{LNP: this def can be simplified by only talking about edges, since edge have a src/tgt node info attached too. beware the current def is incomplete, the nodes produced by me must be the ones matched by mn exactly. Its becoming clear i need to change the defs of node and edge to make these defs more simple. TBD fri.}

A graph pattern can be viewed as a pattern-matching rule on graphs and its precise applicability criteria. It can also be viewed as a sentential form of the language grammar, e.g. \texttt{ScalarVal PlusOp ScalarVal} is a pattern, if the grammar is well formed.

%% Here \texttt{M(leftChild(BinOp)) = \{ scalarValue\}}, in our linear algebra language we would have equivalently \texttt{M(leftChild(BinOp)) = \{ a,b,c,...,0,1\}}. 

\begin{definition}[Graph pattern]
\label{def:graphpattern}
A graph pattern $P$ is an unambiguous structural description of a (sub-)graph $G_P$, which can be deterministically matched in any expression graph $G$. We have $P = \langle G_P, M_n, M_e \rangle$ where for each node $n_i \in N^{G_P}$, $\{n_{match}\} = M_n(n_i)$ returns the set of node values $n_{match}$ accepted to match $n_i$ on a graph $G$. For $n_i,n_j \in N^{G_P}$, $e_i = M_e(n_i, n_j)$ returns the set of edges between $M(n_i)$ and $M(n_j)$ to be matched in $G$. A pattern $G_P$ is matched in $G$ if (a) $\forall n_i \in G_p,~ \exists~ n_m = M(n_i) \in N^G$; (b) $\forall e_i \in E^{G_P}, \exists~ e_{M_n(n_i),M_n(n_j)} = M_e(n_i, n_j) \in E^G$; and (c) $\not \exists e_{M_n(n_i),M_n(n_j)} \in E^G \ne M_e(n_i, n_j)$.
\end{definition}

Note when a graph pattern models a rewrite, $M_n$ and $M_e$ are adjusted accordingly to output the rewrite of a node $n \in N^G$ into its desired value, instead of the set of acceptable nodes from $n \in N^{G_P}$.

\begin{definition}[Axiom of equivalence] An axiom $A$ is a semantics-preserving rewrite rule $G' = A(n,G)$ that can arbitrarily modify a expression graph $G$, and produces another expression graph $G'$ respecting Def.~\ref{def:proggraph} with identical semantics to $G$. We note $A : \langle P_{match}, P_{replace} \rangle$ an axiom, where $P_{match}, P_{replace}$ are graph patterns as per Def.~\ref{def:graphpattern}. The application of axiom $A$ to node $n$ in $G$ is written $A(n,G)$.
\end{definition}

We can compose axioms to form a complex rewrite sequence.

\begin{definition}[Semantics-preserving axiom composition]
\label{def:axiomcompos}
Given a sequence $S:~ A_1(n_1,A_2(n_2,...,A_m(n_m,G)))$ of $m$ axioms applications. It is a semantics-preserving composition if for each $G_j = A_i(n_i,G_i) \in S$, $P_{match}^{A_i}$ succeeds on the subgraph with root $n_i$ in $G_i$, and $G_j$ is obtained by applying $P_{replace}^{A_i}$ to $n_i$.
\end{definition}

\begin{theorem}[Expression graph equivalence]
\label{th:progequiv}
Given a expression $G$. If $G' = S(G)$ such that $S$ is a semantics-preserving sequence as per Def.~\ref{def:axiomcompos}, then $G \equiv G'$, they are equivalent under the axiom system used in $S$.
\end{theorem}

This is a direct consequence of using only semantics-preserving axioms, each rewrite cannot individually alter the semantics, so such incremental composition does not. It leads to the formal problem we are addressing:

\begin{corollary}[Expression graphs equivalence matching]
\label{th:progequivmatching}
Given two expressions $G,G'$. If there exist a semantics-preserving sequence $S$ such that $G' = S(G)$, then $G \equiv G'$.
\end{corollary}

Note here $=$ means complete structural equivalence between the two graphs: they are identical in structure \emph{and} label/node values. Determining $G = G'$ amounts to visiting both graphs simultaneously e.g. in depth-first search from the root to ensure structural equivalence, and also verifying the same node labels appear in both at the same time. This is trivally implemented in linear time in the graph size.

\paragraph*{Language of linear algebra expressions} We have implememented a total of 102 different axioms for our language, made of the multi-type versions of the 13 core restructuring axioms described later in Table~\ref{tab:TransformPct}. They all follow established linear algebra properties. Note different data types have different axioms following typical linear algebra rules, e.g., matrix-multiplication does not commute, but scalar and vector multiplications do. Examples of axioms include $x(yz) \rightarrow (xy)z$, $X-X\rightarrow O$, $-(\vec x - \vec y) \rightarrow \vec y - \vec x$, or $X^{t^t} \rightarrow X$, an exhaustive list is displayed in the Supplementary Material.

In our experiments, we presume matrix and vector dimensions are appropriate for the given operation. Such dimension compatibility
checks are simple to implement by e.g. introducing additional nodes in the prgram representation, but are not considered in our test language.

%% The DNN system we build is able to re-discover these subtleties, as shown in Sec.~\ref{sec:expresults}.

%% This will challenge the network enough to understand 

%% \FIXME{LNP: insert here the table with base axioms. Must carefully think about the syntax. backup is to bring the table from later sec.}

\paragraph*{Examples} We illustrate axiom-based rewrites using axioms presented in later Table~\ref{tab:TransformPct}. Note axiom names follow the structural changes applied. For example, we have $a+b \equiv b+a:~\{a+b\}= Commute(\{+\},\{b+a\})$. $a+b+c \equiv b+c+a:~\{a+b+c\}= Commute(\{+_1\},Commute(\{+_2\},\{b+c+a\})$. Note we refer to different nodes with the same symbol (e.g., $+_2$) subscripting them by their order in a DFS traversal of the expression graph, starting from the unique root. We have $0 \equiv a-a:~\{0\}= Cancel(\{-\},\{a-a\})$. These can be combined in complex paths, e.g., $b+c \equiv c+b+(a-a):~\{b+c\}= Commute(\{+\},Noop(\{+\},Cancel(\{-\},\{c+b+(a-a)\})))$. Such axioms are developed for scalars, matrices and vectors, and include complex rewrites such as distributivity rules and transpositions. A total of 102 axioms are used in our system.

\subsection{Space of Equivalences}

We now define the search space being explored in this work, i.e., the exact space of solutions on which the DNN system formally operates, and that we sample for training.

\begin{definition}[Graph of the space of equivalences]
\label{def:graphofequiv}
Given a language $\mathcal{L}$. The directed graph of equivalences between expressions is $G^{equiv} = \langle N^{equiv}, E^{equiv}\rangle$ such that $\forall l \in \mathcal{L}, n_l \in N^{equiv}$, and $e^{A_i,x}_{n_i,n_j} : n_i \rightarrow n_j \in E^{equiv}$ iff $n_j \equiv A_i(x,n_i)$, $\forall A_i$ in the axiom system and $x$ a position in $n_i$ where $A_i$ is applicable.
\end{definition}

In other words, the graph has one node per possible expression in the language $\mathcal{L}$, and a single axiom application leads to connecting two nodes. We immediately note that $G^{equiv}$ is a (possibly infinite) multigraph, and contains circuits.

\begin{theorem}[Expression equivalence with pathfinding]
Given two expressions $n_i,n_j \in N^{equiv}$. If there is any path from $n_i$ to $n_j$ in $G^{equiv}$, then $n_i \equiv n_j$.
\end{theorem}

The proof is a direct consequence of Def.~\ref{def:graphofequiv}. In this work, we randomly sample this exact graph to learn how to build paths between arbitrary expressions. As it is a multigraph, there will be possibly many different sequences modeled to prove the equivalence between two expressions. It is sufficient to expose one to prove equivalence.

\begin{corollary}[Semantics-preserving rewrite sequence]
Any directed path in $G^{equiv}$ is a semantics-preserving rewrite sequence between the expressions, described by the sequence of axioms and expression position labeling the edges in this path. This sequence forms the proof of equivalence.
\end{corollary}

We believe that ensuring there are possibly (usually) many ways to compute a proof of equivalence in our specific framework is key to enable the DNN approach to learn automatically the pathfinding algorithm for building such proofs. Other more compact representations of this space of equivalences are clearly possible, including by folding nodes in the equivalence graph for structurally-similar expressions and folding equivalent paths between nodes. When building e.g. a deterministic algorithm for pathfinding, such space size reduction would bring complexity benefits \cite{kaplan1969regular,barthou2002}. We believe that for the efficient deployment of graph-to-sequence systems, exposing significant redundancy in the space facilitates the learning process. We also alleviate the need to reason on the properties of this space to find an efficient traversal heuristic.

\section{Samples Generation}
\label{sec:samplegen}
%% \section{Samples Generation}
%% \label{sec:samplegen}
%% \input{samplegen}

The careful creation of our training dataset is key: as we let the DNN learn \emph{by example only} what the axioms are and when they are applicable in the structure of a program, we must carefully sample the space of equivalences to ensure appropriate distributions of the examples. We produce a final dataset of tuples $(P1,P2,S)$, a pair of input programs and a possible rewrite rule sequence that proves the pair equivalent. Duplicates are removed such that all samples have a unique $P1$. From this dataset, we create 1,000,000 training samples, 10,000 validation samples, and 10,000 test samples. 
We outline below its generation principles; extensive details and the algorithms used are presented in section~\ref{sec:appendix:genexamples}.

\paragraph*{Random sample generation}
Deep learning typically requires large training sets to be effectively deployed, hence we developed a process to automate sample generation. We specifically use randomized program generation algorithms that are inspired by a given language grammar. By randomly choosing between production rulse, one can build random parse trees by simply iterating the grammar. The leaves obtained will form a sentence accepted by the language, i.e., a program \cite{bielik16}.
We limit to programs of 50 nodes in the program tree (or AST), with a maximal tree depth of 7.
%Up to our size limits of 50 AST nodes and AST depth of 7, 
We assert that our random production rule procedure has a non-zero probability of producing any program allowed by the grammar for our datasets.

We produce equivalent program samples by pseudo-randomly applying axioms on one randomly generated program to produce a rewrite sequence and the associated equivalent program. Given a randomly selected node in the program graph, our process checks which axiom(s) can be applied. E.g., the $+_m$ operator may have the Commute axiom category applied, or it may have the Transpose axiom category applied, which affects the operator's children.

\paragraph*{Final experimental dataset: AxiomStep10}
%As shown in Figure~\ref{fig:Full} and detailed further in Section~\ref{sec:progequivdnn}, our network is trained to produce one axiom step at a time. 
%As shown in Figure~\ref{fig:Full} and detailed further in Section~\ref{sec:progequivdnn}, 
To train our network to produce one axiom step at a time, as described in Sec.~\ref{sec:motivation}, AxiomStep10 has a single axiom in each output sequence $S$.
For a complete proof $S:A_1(A_2(...)$ in a $(P1,P2,S)$ we generated made of N axioms, we then create N training examples for the network: $(P1,P2,A_N)$ the first intermediate step by applying the first axiom, then $(A_N(P1),P2,A_{N-1})$, etc.
%The first training sample has the original program; later samples have intermediate programs; all samples include the target program. 
%Hence, given a set of program pairs and axiomatic proofs, for a proof of length N we create N training examples for the network. The first training sample has the original program; later samples have intermediate programs; all samples include the target program.
%Hence, given a set of program pairs and axiomatic proofs, for a proof of length N we create N training examples for the network. The first training sample has the original program; later samples have intermediate programs; all samples include the target program.
%Given our maximum axiom distance between programs is 10, we call this axiom step dataset AxiomStep10. Test samples only have the original and target program and the network proposes axioms which create intermediate programs towards the proof. 
We limit proof length to 10 axioms in our experiments (hence AxiomStep10). Test samples only have the original and target program and the network proposes axioms which create intermediate programs towards the proof, fed back to the system. 

\begin{table}[h!tb]
\caption{Distribution for the 14 axiom categories in AxiomStep10 test set. Considering scalars (a, b, ...), vectors ($\vec v$,$\vec w$, ...) and matrices (A, B, ...) types combinations, 147 distinct axioms are represented.\vspace{-.3cm}}
  \label{tab:TransformPct}
  \small
  \setlength\tabcolsep{2pt}
  \centering
  \begin{tabular}{@{}llr@{}}
    \toprule
 Axiom Category & Example axiom(s) & Samples with \\
    \midrule
 Cancel & (A-A)$\rightarrow$O,(b/b)$\rightarrow$1 & 13.8\% \\
 NeutralOp & ($\vec v$ - $\vec o$) $\rightarrow$ $\vec v$ & 40.0\% \\
 DoubleOp & $A^{t^t} \rightarrow A$, 1/1/x$\rightarrow$x & 7.3\% \\
 AbsorbOp & (A*O)$\rightarrow$O, (b*0)$\rightarrow$0 & 30.3\% \\
 Commute & (a + b) $\rightarrow$ (b + a) & 48.6\% \\
 DistributeLeft & (a + b)c $\rightarrow$ ac + bc & 36.3\% \\
 DistributeRight & a(b + c) $\rightarrow$ ab + ac & 27.8\% \\
 FactorLeft & ab + ac $\rightarrow$ a(b+c) & 6.1\% \\
 FactorRight & ac + bc $\rightarrow$ (a+b)c & 9.0\% \\
 AssociativeLeft & a(bc) $\rightarrow$ (ab)c & 46.3\%  \\
 AssociativeRight & (ab)c $\rightarrow$ a(bc) & 43.1\% \\
 FlipLeft & -($\vec v$ - $\vec w$) $\rightarrow$ $\vec w-\vec v$ & 8.4\% \\
 FlipRight & a/(b/c) $\rightarrow$ a(c/b) & 26.1\% \\
 Transpose & $(AB)^{t} \rightarrow B^{t}A^{t}$,  & 11.1\% \\
     \bottomrule
  \end{tabular}
 %%  \begin{tabular}{@{}llclllc@{}}
 %%    \toprule
 %% Axiom Category & Example axiom(s) & Samples with & & Axiom Category & Example axiom(s) & Samples with \\
 %%    \midrule
 %% Cancel & (A-A)$\rightarrow$O,(b/b)$\rightarrow$1 & 13.8\% & &  NeutralOp & ($\vec v$ - $\vec o$) $\rightarrow$ $\vec v$ & 40.0\% \\
 %% DoubleOp & $A^{t^t} \rightarrow A$, 1/1/x$\rightarrow$x & 7.3\% & & AbsorbOp & (A*O)$\rightarrow$O, (b*0)$\rightarrow$0 & 30.3\% \\
 %% Commute & (a + b) $\rightarrow$ (b + a) & 48.6\% & &  DistributeLeft & (a + b)c $\rightarrow$ ac + bc & 36.3\% \\
 %% DistributeRight & a(b + c) $\rightarrow$ ab + ac & 27.8\% & & FactorLeft & ab + ac $\rightarrow$ a(b+c) & 6.1\% \\
 %% FactorRight & ac + bc $\rightarrow$ (a+b)c & 9.0\% & & AssociativeLeft & a(bc) $\rightarrow$ (ab)c & 46.3\%  \\
 %% AssociativeRight & (ab)c $\rightarrow$ a(bc) & 43.1\% & & FlipLeft & -($\vec v$ - $\vec w$) $\rightarrow$ $\vec w-\vec v$ & 8.4\% \\
 %% FlipRight & a/(b/c) $\rightarrow$ a(c/b) & 26.1\% & & Transpose & $(AB)^{t} \rightarrow B^{t}A^{t}$,  & 11.1\% \\
 %%     \bottomrule
 %%  \end{tabular}
\end{table}
\vspace{-.3cm}

\paragraph*{Datasets to study generalizability and robustness}
In order to study our model's ability to generalize, we have created alternate datasets on which to train and test models which are summarized in table~\ref{tab:Datasets}.
\emph{WholeProof10} will help us 
%explore how valuable the axiom step output of our model is. 
contrast learning approaches.
%This dataset includes the whole proof as the target output instead of a single axiom step. 
This dataset has the complete proof sequence $S$ made of $N\ge 1$ axioms as reference output for a program pair, while for AxiomStep10, $N=1$.
Models trained on WholeProofX must maintain internal state representing the graph transformations that the axioms create. They are not "iterative": a single inference is expected to produce the complete proof; in contrast to AxiomStep10 for which a single axiom of the sequence is produced at each inference step. Training long output sequences can benefit from complex training approaches such as Professor forcing \cite{Lamb16}, but we will show that our AxiomStep10 model generalizes well with our sequence model training approach.
%, requiring iterative processing as shown in Fig.~\ref{fig:Full}.
%%% LNP to Steve: I could not understand what you meant here with the dataset created with a preorder pass through the ast. I don't think any reader can easily get it in a single sentence. But the budget is 15 words or less... I commented out.
%In deference to this complexity, this dataset is created with a single preorder pass through the nodes of the AST, providing a more consistent rewrite sequence for training. 

\begin{table}[h!tb]
\vspace{-.3cm}
  \caption{Datasets used for studies in experiments.\vspace{-.3cm}}
 % \caption{Datasets used for studies in experiments. AxiomStep5 and WholeProof5 allow for training models on less general datasets to explore generalization to the testsets with proof lengths up to 10.}
    \label{tab:Datasets}
  \small
  \centering
  \begin{tabular}{@{}lcccl@{}}
    \toprule
 Dataset & AST depth & AST \#nodes & Proof length & Iterative \\
    \midrule
 AxiomStep10 & 2-7 & 2-50 & 1-10 & Yes\\
 AxiomStep5 & 2-6 & 2-25 & 1-5 & Yes\\
 WholeProof10 & 2-7 & 2-50 & 1-10 & No\\
 WholeProof5 & 2-6 & 2-25 & 1-5 & No\\
    \bottomrule
  \end{tabular}
\end{table}

\paragraph*{Complexity of equivalence space}

\begin{figure}[h!tb]
\vspace{-.3cm}
 \includegraphics[width=7cm]{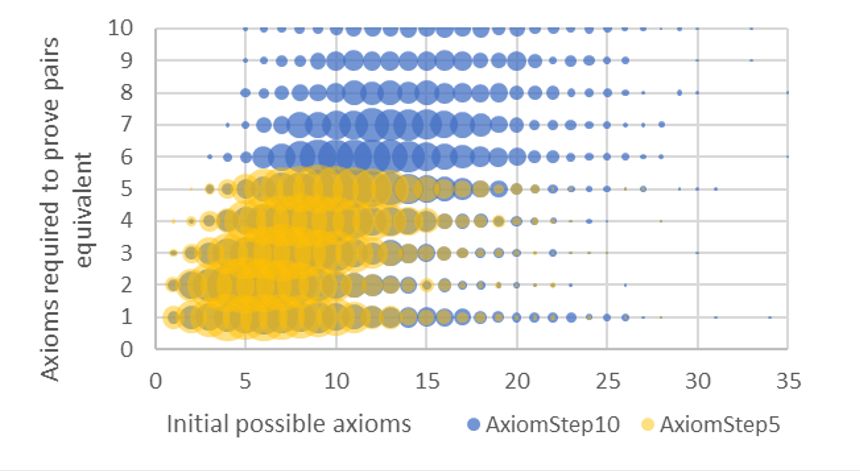}
 \includegraphics[width=7cm]{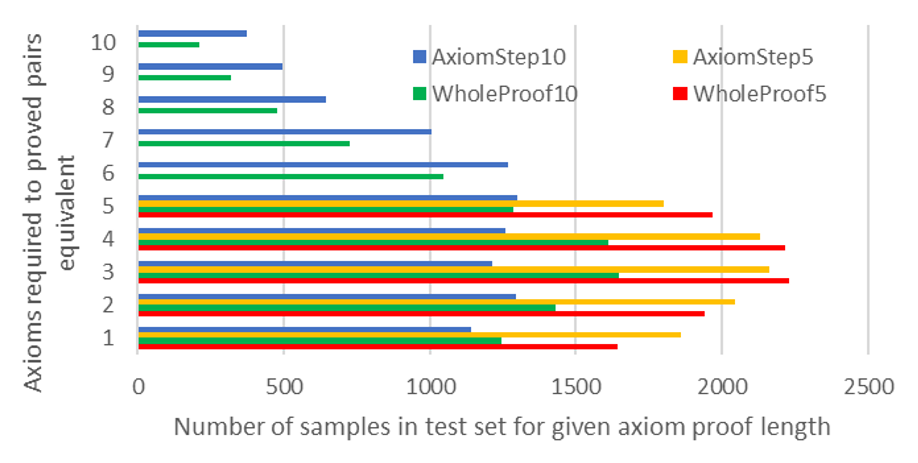}
\vspace{-.6cm}
\caption{Distribution of axiom possibilities and proof complexity for test datasets.}
\label{fig:Datadist}
\vspace{-.5cm}
\end{figure}

Figure~\ref{fig:Datadist} provides a view of the complexity of the
equivalence problem we tackle. The distribution of the dataset per
proof length is displayed in the right chart; the left chart shows by
size of bubble the number of test samples with a given number
of \emph{semantics-preserving} axioms that may be implemented as the
first step of the proof and the proof length needed.

There is a large number of proofs possible in our system, as detailed
in Appendix~\ref{subsec:complexityofprovingequiv}. For example, for
proofs of length 5, about 340,000 proofs made only of legal
applications of axioms can be performed on the average sample in our
dataset. Since many programs have multiple possible proofs, about
10,000 different programs can be produced, only one of which is the
target to prove, i.e., randomly drawing a valid 5 axiom proof on a
program known to be 5 axiom steps from the target has roughly a 1 in
10,000 chance of being a correct proof of equivalence between the two
programs.

\section{Deep Neural Networks for Program Equivalence}
\label{sec:progequivdnn}
Fig.~\ref{fig:Full} overviews the entire system architecture including sample generation, the \texttt{pe-graph2axiom} network, and the rewrite checker. Key design decisions are presented below.

%\FIXME{LNP move System components paragraph to section 2}
%\paragraph*{System components}
%The system in Fig.~\ref{fig:Full}  is composed of the following blocks. 
%\emph{Node initialization} is the process in which the program is used to initialize the node data structures used by the graph neural network.
%\emph{Graph neural network} refers to a neural network that has weights which allow it to learn interrelations between network nodes based on edge connections.
%\emph{Global attention} \cite{luong15} when used with a graph neural network allows the decoder to pay attention to certain nodes in the graph as it creates each token in the output sequence.
%\emph{Token embedding} is a neural network layer in which tokens are assigned a learnable multidimensional embedding vector \cite{Mikolov13}.
%\emph{LSTM 2 layers} is referring to 2 layers of Long Short Term Memory (LSTM) neurons, each layer has 256 neurons, which support sequence generation.
%\emph{Token generator} is the final output portion of the network. It learns to output the tokens based on the current LSTM hidden states and the global attention from the graph neural network. As each token is output, it feeds back into the LSTM layer through the embedding layer to affect its next state.

\paragraph*{Graph neural network}
The sample generation discussed in section \ref{sec:samplegen}
provides input to the \textsf{Node Initialization} module in
Fig.~\ref{fig:Full} to create the initial state of our graph neural
network \citep{Beck18}. For each node in the program graph, a node will be initialized in our
graph neural network with a value that encodes the AST level and language token of the program node.
To interconnect the edges we support 9 edge types and their reverse edges which allows information to move in any direction necessary: 1) left child of binary
op, 2) right child of binary op, 3) child of unary op, 4) root node to
program 1, 5) root node to program 2, 6-9) there are 4 edge
types for the node grandchildren (LL, LR, RL, RR). The node states and edge adjacency matrix represent the initial graph neural network state. 

After initialization, the graph neural network iterates 10 times in order to convert the initial node state into the embeddings needed for rewrite rule generation.
Given an initial hidden state for node $n$ of $x_n(0)$, $x_n(t+1)$ is computed with a learnable function $f$ which combines the current hidden state $x_n(0)$, the edge types $l_{in[n]}$ of edges entering node $n$, the edge types $l_{out[n]}$ of edges exiting node $n$, and the hidden states $x_{ne[n]}$ of the neighbors of node $n$: $ x_n(t+1) = f(x_n(t),l_{in[n]},x_{ne[n]}(t),l_{out[n]}) $.

Each edge type has a different weight matrix for learning,
allowing aggregation of information into a given node related to its function in the full graph of the program. The root node's initial state along with the edge types connecting it to the program graph trees allow it to aggregate and transfer specific information regarding rewrite rules as demonstrated by our experimental results. This is a novel feature of our network not used in prior work with GNNs on program analysis \citep{AllamanisICLR18,Xu17}.

% \paragraph*{Graph neural network output to decoder} Fig.~\ref{fig:Network} shows two ways that the final node values for the graph are used by the decoder to create rewrite rules. First, the final root node value $x_{root}(10)$ is fed through a learnable bridge function to initialize the LSTMs of the decoder. In this way, the aggregated information of the 2 programs seeds the generation of rewrite rules. The LSTMs update as each output token $y_{j}$ is generated with a learnable function based on the current decoder hidden state $h_{j}^d$ at decoder step $j$ and the previous output token $y_{j-1}$ \citep{Chen19}. Second, all nodes in the graph can be used by an attention layer \citep{Bahdanau14}. The attention layer creates a context vector $c_j$ which can be used by a learnable function $g$ when computing the probability for generating the $j$th output token $P(y_{j})$: $
\paragraph*{Graph neural network output to decoder} After stepping the GGNN, the final node values are used by the decoder in two ways to create rewrite rules. First, the final root node value $x_{root}(10)$ is fed through a learnable bridge function to initialize the LSTMs of the decoder. In this way, the aggregated information of the 2 programs seeds the generation of rewrite rules. The LSTMs update as each output token $y_{j}$ is generated with a learnable function based on the current decoder hidden state $h_{j}^d$ at decoder step $j$ and the previous output token $y_{j-1}$ \citep{Chen19}. Second, all nodes in the graph can be used by an attention layer \citep{Bahdanau14}. The attention layer creates a context vector $c_j$ which can be used by a learnable function $g$ when computing the probability for generating the $j$th output token $P(y_{j})$: $
%
%\begin{equation}
%\label{eq:attntokenprob}\nolabel
P(y_{j} \mid y_{j-1},y_{j-2},...,y_{0},c_j) = g(h_{j}^d, y_{j-1}, c_j)
%\end{equation}
$. Because \texttt{pe-graph2axiom} has a robust output verification, we make use of beam search to track up to 10 likely candidates for proofs of equivalence.

By using the root node only for seeding the initial hidden state $h_{0}^d$ of the decoder, the weights associated with its connections to the program graphs for $P1$ and $P2$ learn to represent the information necessary for the rewrite rule sequence. In parallel, after the graph neural network iterations complete, the final embedding for all the nodes in the graphs for $P1$ and $P2$ are only used by the attention network, so their final embedding represents information useful during rewrite rule generation. 

% Moved to appendix
% \begin{figure}[h!tb]
% \vspace{-.5cm}
% \centering
% \includegraphics[width=0.8\textwidth]{images/GGNNdetail.png}
% \caption{Graph-to-sequence neural network data flow details.}
% \label{fig:Network}
% \vspace{-.5cm}
% \end{figure}

%\paragraph*{Intermediate program generation} 
\paragraph*{Intermediate program generation}
\texttt{pe-graph2axiom} applies the axiom and program node chosen by the neural network token generator to the input program to create an intermediate program $P'$ on the path from $P1$ to $P2$. If this program is equal to $P2$, then our axiom path is complete, otherwise the new pair $P',P2$ is inferred to determine the next axiom step.

% Moved to appendix; beam search mentioned near decoder now above.
%\paragraph*{Beam search}
%A typical approach when using sequence-to-sequence systems is to
%enable \emph{beam search}, the process of asking for multiple answers to the same question to the network. It is particularly relevant when creating outputs which can be automatically checked \cite{Chen19,ahmed18}. Given the stochastic nature of a generation model, a beam width of $n$ can be thought of as creating the $n$ most likely sequences given the training data the model as learned on and the input provided. Our model builds on the beam search provided by OpenNMT to create a 'system beam search' as it searches for a successful proof using intermediate programs, details of this process are provided in the Supplementary Materials. We evaluate in
%Sec.~\ref{sec:expresults} beam sizes ranging from 1 to 10, 
%showing higher success with larger beams.

\paragraph*{Incremental versus non-incremental sequence production}
The models we train on the AxiomStep5, WholeProof10, and WholeProof5 datasets have the same neural network hyperparemeters as the AxiomStep10 data model. However, the models for WholeProof10 and WholeProof5 are trained to output the entire sequence of axioms needed to prove the 2 programs identical, hence these models do not make use of the intermediate program generation and instead have a component which checks whether the full sequence of axioms legally transforms $P1$ into $P2$.
We encode the path to the AST node an which to apply an axiom using 'left' and 'right' tokens which specify the path from the current program root node. This encoding is sufficient for the iterative model and necessary to allow the non-iterative model to identify nodes which may not have been in the initial AST for $P1$. 
%%% LNP: temporary removal to see alignment for sec 6.
The non-iterative models must learn a representation in the LSTM network to allow them to track AST transformations as they are generated.

\section{Experimental Results}
\label{sec:expresults}
%% \section{Experimental Results}
%% \label{sec:expresults}
%% \input{expresults}

We now present extensive experimental results, and compare the quality of several neural network approaches to address the problem of program equivalence. We have proceeded incrementally for fine-tuning the final system design, and report on several of these design points below.

%After experiments to validate our model architecture and hyperparameters, 
We focus our experiments below on 4 key questions: 1) Is performance related to input program size? 2) Is performance related to proof length? 3) Is the incremental, per-axiom approach more generalizable than producing the full sequence in a single inference step? And 4) Is performance consistent across a range of datasets, including human-written examples?

%Does our model generalize to unseen input programs well? 2) Does it generalize to 
%2) How broadly does our training data cover the space of equivalent programs? 
%3) Will our model perform well on human problems from the community? 

\paragraph*{Implementation setup}
\label{sec:expresults:setup}
We developed the neural network system presented in the OpenNMT-py system \citep{opennmt}, adding on a new encoder based on a prior implementation of gated graph neural networks \citep{Li16}. For our training and evaluation experiments, we use systems with Intel Xeon 3.6GHz CPUs and 6GB GeForce GTX 1060 GPUs. 
During training, we save a model snapshot every 50,000 iterations and score the accuracy the model achieved on the validation dataset. Graphs showing that validation accuracy plateaus at 200,000 to 300,000 iterations are provided in section~\ref{sec:suppl:additionalresults}. We run each model twice and evaluate the test set using the saved model which achieved the highest validation score.

\paragraph*{Evaluation procedure and neural network alternatives}
% During training and validation we are selecting models based on the per-token accuracy of the model with the target axiom and axiom position. However for our test dataset evaluation, we instead report the accuracy of the model to output a correct rewrite sequence with beam sizes of 1,2,5, or 10. 
%The benefits of key components of our neural network model are demonstrated in table~\ref{tab:metamodel}. The bidirectional RNN model is similar to state-of-the-art sequence-to-sequence models used for program repair \cite{Chen19}. The input to this model in a parenthesized string representing the equation to learn. Results with that model demonstrate the benefit of analyzing the input program as a graph, where the model resources (the weights of the edge matrices) can more directly model the salient features of the 2 programs. The relative results for the graph-to-sequence model without attention show the benefit of providing the node information during the axiom generation process, although since the AxiomStep10 system is only producing a single axiom at a specific location, the performance without attention is not significantly worse than \texttt{pe-graph2axiom}.
The benefits of key components of our neural network model are studied in table~\ref{tab:metamodel}. The bidirectional RNN model is similar to state-of-the-art sequence-to-sequence models used for program repair \citep{Chen19}. The  results for the graph-to-sequence model without attention show the benefit of providing the node information during the axiom generation process.

\begin{table}[h!tb]
  \vspace{-.3cm}
  \caption{\texttt{pe-graph2axiom} mini ablation study.\vspace{-.3cm} }
  %  \caption{\texttt{pe-graph2axiom} is able to generate equivalence proofs for 93\% of the AxiomStep10 tests. Other model variants produce a lower percentage of correct proofs.}
  
  \label{tab:metamodel}
  \small
  \centering
  \begin{tabular}{@{}lrrrrr@{}}
    \toprule
    & & \multicolumn{4}{c}{Beam width} \\
    Model description & & 1 & 2 & 5 & 10   \\
    \cmidrule{1-1} \cmidrule{3-6}
    Bidirectional RNN seq-to-seq with attention & & 48 & 62 & 71 & 75 \\
    Graph-to-sequence w/o attention & & 73 & 81 & 87 & 90 \\
    \texttt{pe-graph2axiom} model & & 76 & 84 & 90 & \textbf{93} \\
    \bottomrule
  \end{tabular}
\end{table}

Our final design was influenced by explorations we performed on varied models, datasets, and hyperparameters such as LSTM layers and graph neural network parameters. In relation to the model's ability to learn a representation of the proof sequence, we note that our GGNN initialization using the root node connection to the decoder outperforms the embedding learned by a bidirectional RNN model. Also, we found that averaging the embedding of all graph nodes had about 10\% lower accuracy than using the more specific root node information. Numerous additional results are reported in Suppl. material~\ref{sec:suppl:additionalresults}. 

% Model generalizing across varying dataset distributions
\paragraph*{Generalizing across different datasets}
%\FIXME{Can we cut these paragraphs in half somehow?}
We specifically look at the generalization potential for our models by studying their success rate as a function of the input program complexity, represented as the AST depth, in Table~\ref{tab:newtreedepth}, and as a function of the output complexity, represented by the proof length in Table~\ref{tab:modeltest}, all using a beam size of 10. We designed our datasets in Sec.~\ref{sec:samplegen} to study how well \texttt{pe-graph2axiom} generalizes and to assess we are not overfitting on training data. Extensive in-depth additional experimental results are presented in Suppl. Material~\ref{sec:suppl:additionalresults}, we summarize key results only below.

\begin{table}[h!tb]
\vspace{-.3cm}
\caption{\label{tab:newtreedepth}Performance vs. AST size: counts and percentage pass rates.\vspace{-.3cm}}
%  \caption{Study on generalizing to larger AST sizes. When considering the 10,000 tests from each of the AxiomStep5 set (AS5) and AxiomStep10 (AS10), columns show sample counts and percent pass rates for P1 having increasing AST depths. The model trained with the AxiomStep5 dataset had no training examples with AST depths of 7 yet it scales well to these more complex problems.}
  \scriptsize
  \centering
%\vspace{-.5cm}
\begin{tabular}{@{}rrrrrrrrrr@{}}
    \toprule
    & & \multicolumn{2}{c}{Testset} & & \multicolumn{2}{c}{Model trained} & & \multicolumn{2}{c}{Model trained}  \\
    & & \multicolumn{2}{c}{Sample Count} & & \multicolumn{2}{c}{on AxiomStep5} & & \multicolumn{2}{c}{on AxiomStep10}  \\
    \cmidrule{3-4} \cmidrule{6-7} \cmidrule{9-10}
    AST depth & & AS5 & AS10 & &  AS5 & AS10 & & AS5 & AS10 \\
    \midrule
%    2 & & 5 & 3 & & 100 & 100 & & 100 & 100 \\
%    3 & & 306 & 133 & & 100 & 100 & & 100 & 100 \\
%    4 & & 1489 & 577 & & 100 & 99 & & 99 & 99 \\
%    5 & & 4744 & 1844 & & 99 & 94 & & 98 & 95 \\
%    6 & & 3456 & 4308 & & 98 & 90 & & 98 & 93 \\
  2-6 & & 10000 & 6865 & & 99 & 93 & & 99 & 94 \\
    7 & & 0 & 3135 & & n/a & 86 & & n/a & 92 \\
    All & & 10000 & 10000 & & 99 & 90 & & 99 & \textbf{93} \\
    \bottomrule
  \end{tabular}
  \vspace{-.3cm}
\end{table}

Table~\ref{tab:newtreedepth} illustrates the ability of a model trained on AxiomStep5 (i.e., limited to proofs of length 5) to perform well when evaluated on the more complex AxiomStep10, which includes proofs of unseen length of up to 10. The robustness to the input program complexity is illustrated with the 86\% pass rate on AST depth 7, for the model trained on AxiomStep5 which never saw programs of depth 7 during training.

\begin{table*}[h!tb]
\vspace{-.3cm}
  \caption{Performance vs. proof length: percentage pass rates.\vspace{-.3cm}}
 %   \caption{Study on generalizing to longer proofs. Percentage pass rates for equivalence proofs of increasing axiom counts when testing each of 4 datasets on models trained using each of 4 datasets.}
  \label{tab:modeltest}
  \scriptsize
  \centering
  \setlength\tabcolsep{2pt}
  \begin{tabular}{@{}r|rrrrrrrrrrrrrrrrrrrr@{}}
    \toprule
    Axiom & & \multicolumn{4}{c}{Model trained on} & & \multicolumn{4}{c}{Model trained on} & & \multicolumn{4}{c}{Model trained on} & & \multicolumn{4}{c}{Model trained on}   \\
    Count in & & \multicolumn{4}{c}{WholeProof5 (WP5)} & & \multicolumn{4}{c}{WholeProof10 (WP10)} & & \multicolumn{4}{c}{AxiomStep5 (AS5)} & & \multicolumn{4}{c}{AxiomStep10 (AS10)}   \\
    \cmidrule{3-6} \cmidrule{8-11} \cmidrule{13-16} \cmidrule{18-21}
    Proof & & WP5 & \scriptsize{WP10} & AS5 & \scriptsize{AS10} & & WP5 & \scriptsize{WP10} & AS5 & \scriptsize{AS10} & & WP5 & \scriptsize{WP10} & AS5 & \scriptsize{AS10} & & WP5 & \scriptsize{WP10} & AS5 & \scriptsize{AS10} \\
    \midrule
    1-5 & & 95 & 89 & 44 & 44 & & 94 & 93 & 44 & 44 & & 99 & 97 & 99 & 98 & & 99 & 98 & 99 & \textbf{98} \\
    6 & &   & 14 &   & 4 & &   & 72 &   & 5 & &   & 81 &   & 88 & &   & 90 &   & \textbf{93} \\
    7 & &   & 0 &   & 1 & &   & 63 &   & 2 & &   & 67 &   & 81 & &   & 83 &   & \textbf{87} \\
    8 & &   & 0 &   & 0 & &   & 54 &   & 1 & &   & 54 &   & 75 & &   & 73 &   & \textbf{82} \\
    9 & &   & 0 &   & 0 & &   & 47 &   & 0 & &   & 35 &   & 64 & &   & 63 &   & \textbf{74} \\
    10 & &   & 0 &   & 0 & &   & 34 &   & 0 & &   & 24 &   & 57 & &   & 46 &   & \textbf{66} \\
    All & & 95 & 66 & 44 & 27 & & 94 & 84 & 44 & 27 & & 99 & 87 & 99 & 90 & & 99 & 93 & 99 & \textbf{93} \\
    \bottomrule
  \end{tabular}
  \vspace{-.3cm}
\end{table*}

Table~\ref{tab:modeltest} compares the results of our 4 models, each trained on one of our 4 datasets, and evaluated with the test set of all 4 datasets. The models all have identical hypermeter settings.
%Aside from training on the dataset distributions shown in Figure~\ref{fig:Datadist}, 
%
%First, in relation to dataset breadth, the model trained on WholeProof10 does reasonably well (84\% success) on proving test programs from its trained distribution. On the other hand, 
%
We observe the inability of models trained to output the whole proof to generalize to proofs of higher length (WP5 model on AS10/WP10), with near zero success rate. However, per-axiom models (AS5 and AS10) show potential for generalization to proof length: AS5 model performs well when evaluated on AS10, showing the ability to produce proofs of length/complexity unseen in training. Overall, the success rate degrades gracefully with proof length, bottoming at 66\% for AS10 for proofs of length 10.
%%% LNP: Commenting this out. See the above new text instead.
%AxiomStep10, while training on our broadest dataset in which axioms can be applied to nodes repeatedly and in variable order, achieves a 93\% average success rate. 72\% of the proofs of length 6 from the WholeProof10 testset were solved by the model trained on WholeProof10, but only 5\% of such proofs from AxiomStep10 were, suggesting the method of generating AxiomStep pairs covers the problem space more thoroughly. 
%%% LNP: Better to say /why/ we observe this, not useful to state what we observe.
%In contrast, the model trained on AxiomStep10 data actually performs better on WholeProof10 data than its own model did and its overall score on that test set matches its score on its own dataset.

%%%%%%%%%%

%%%%%%%%%%%%%%%%%%%%%%%%%%%%%%%%%%%%
%%%%%%%%%%%%%%%%%%%%%%%%%%%%%%%%%%%%
\begin{table*}[h]
  \scriptsize
{\begin{tabular}{ |p{0.4cm}|p{6cm}||p{0.4cm}|p{0.4cm}|p{0.4cm}|p{0.7cm}|p{0.7cm}||p{0.7cm}|p{1cm}||p{0.7cm}|p{0.7cm}| }
 \hline
 ID & Description & \rotatebox{90}{\# Operators} & \rotatebox{90}{\# Axioms} & \rotatebox{90}{\# Operands} & \rotatebox{90}{Program length} & \rotatebox{90}{Rewrite rules length} & \rotatebox{90}{\parbox{2.9cm}{Graph2seq (G2S) \\ or seq2seq (S2S)}} & \rotatebox{90}{Training set size} & \rotatebox{90}{\parbox{2.9cm}{Percent matching \\ with beam width 1}} & \rotatebox{90}{\parbox{2.9cm}{Percent matching \\ with beam width 10}} \\
 \hline
 1 & Rewrite sequence is only single Commute, uses sequence-to-sequence model & 2 & 1 & 10 & 3-19 & 1-5 & S2S & 80,000 & 90.0\% & 96.2\% \\
 \hline
 2 & Rewrite sequence is exactly 2 Commutes, uses sequence-to-sequence model & 2 & 1 & 10 & 5-24 & 3-10 & S2S & 80,000 & 80.3\% & 96.5\% \\
 \hline
 3 & Rewrite sequence exactly 2 Commutes & 2 & 1 & 10 & 5-24 & 3-10 & G2S & 80,000 & 98.9\% & 99.8\% \\
 \hline
 4 & Rewrite sequence exactly 3 Commutes & 2 & 1 & 10 & 7-45 & 5-15 & G2S & 80,000 & 91.4\% & 99.0\% \\
 \hline
 5 & Rewrite sequence 1 to 3 Commutes & 2 & 1 & 10 & 3-45 & 1-15 & G2S & 180,000 & 97.1\% & 99.2\% \\
 \hline
 7 & Commute, Noop, Cancel, Distribute Left, Distribute Right & 4 & 5 & 12 & 3-45 & 1-15 & G2S & 180,000 & 93.1\% & 97.4\% \\
 \hline
 8 & Scalars, Vectors, and Matrixes & 16 & 5 & 20 & 3-30 & 1-25 & G2S & 250,000 & 88.3\% & 95.6\% \\
 \hline
 9 & 13 Axioms & 16 & 13 & 20 & 3-30 & 1-25 & G2S & 400,000 & 85.5\% & 95.5\% \\
 \hline
 10 & Rewrite sequence or Not\_equal & 16 & 13 & 20 & 3-30 & 1-25 & G2S & 500,000 & 79.8\% & 93.8\% \\
 \hline
 11 & Test sequence-to-sequence & 16 & 13 & 20 & 3-30 & 1-25 & S2S & 400,000 & 59.8\% & 81.1\% \\
 \hline
 12 & Add loop axioms & 18 & 15 & 20 & 3-30 & 1-25 & G2S & 400,000 & 83.8\% & 94.7\% \\
 \hline
\end{tabular}}
\caption{Results for various language complexities studied, on non-incremental models (WholeProof).}
\label{tab:ResultLang}
\vspace{-.8cm}
\end{table*}

%%%%%%%%%%%%%%%%%%%%%%%%%%%%%%%%%%%%
%%%%%%%%%%%%%%%%%%%%%%%%%%%%%%%%%%%%

\subsection{WholeProof Models: Language Complexity and Performance}

%% As discussed in later Sec.~\ref{sec:expresults:additionalresults}, we iterated numerous possible designs and approaches to figure out the best-working system for this network. In particular, we evaluated simpler approaches before reaching the complexity of our fina design, to ensure a more complex approach was needed.

Table~\ref{tab:ResultLang} shows the result of 12 different experiments and designs specifically for the WholeProof5 models. In particular, we incrementally increase the problem complexity from rows 1 to 10, increasing the number of \textsf{Operators} that can be used in any input program, of \textsf{Axioms} used in the rewrite sequence, of \textsf{Operands} in any input program, of the maximal number of nodes in an input program graph (the \textsf{Program length}, directly influencing the size of the graph network), and  the \textsf{Rewrite rule length}, which contains the description of paths from the root node to reach the position of application of an axiom, this is directly related to the maximal graph height, itself determined by the maximal program size. Details on each row are provided in Supplementary Material.

We specifically compare against a sequence-to-sequence (S2S) approach, to quantify the gains brought by employing graph-to-sequence (G2S). When the space is small enough, S2S still performs well, especially using aggressive beam search. We recall that by design of our system testing the correctness of one sequence is trivial and deterministic, so one can easily use large beam sizes without any correctness impact nor major performance penalty during inference. For example, inference of beam 1 is about 15ms for our most complex networks, but beam 10 only takes 16ms. Checking correctness is $<<$ 1ms.

Contrasting rows 2 and 3 displays the merits of the G2S approach for our problem: on this simple problem, in fact G2S gets near-perfect accuracy already. Progressively increasing the complexity of the search space, till row 9 and 10, displays a slow but steady decrease in quality, while still maintaining excellent scores near or above 95\% with beam 10. To reassess the limits of a sequence-to-sequence approach, row 9 and 11 can be constrasted: they operate on the same search space, but S2S peaks at 81\% accuracy, while G2S reaches 95\%.

Row 10 displays the result when learning using also samples of non-equivalent programs, using the ``empty path'' symbol Not\_equal. We evaluated this system to measure the impact of training on only equivalent programs vs. also sampling pairs of unconnected nodes in the equivalences graph. We recall that by design, if no rewrite rule produced is verified as correct, our system  outputs the programs are not equivalent. In other words, whichever the sequence(s) produced by the network, if the two input programs are non-equivalent, the system will \emph{always} output they are not equivalent: no equivalence sequence produced can be verified as correct. So training on only equivalent programs is clearly sensible for such system; furthermore as shown in row 10 vs. 9, even increasing the training set size, training using non-equivalent programs seem to lower the performance slightly.

%% Our best result (golden model) with the full language has 9545/10000 exact
%% matches with beam width 10, and 9623/10000 correct
%% proofs of equivalence (i.e., 78 of the 455 cases without an exact match
%% still have a legal rewrite rule sequence produced). 

%%%%%%%%%%5

\paragraph*{Human written test expressions from Khan academy exercises}
Unfortunately there is a dearth of existing large reference datasets for equivalence of linear algebra expressions, which justified our careful dataset creation approach in Sec.~\ref{sec:samplegen} and their upcoming public release. However numerous math exercises involve exactly this problem, and can provide small but human-written datasets. We solve all of the matrix expression equivalence programs from 2 relevant Khan academy modules designed to test student's knowledge of matrix algebra \citep{Khan20}. 
%
%In order to further validate that our model has generalized to real-world problems, we solve all of the matrix expression equivalence programs from 2 relevant Khan academy modules designed to test student's knowledge of matrix algebra \cite{Khan20}. 
%
%We take all 15 equivalent program pairs from these problems and 5 non-equivalent programs. %Indeed, none of our systems claim to find a proof that the non-equivalent programs were equivalent. 
Our AxiomStep10 model is able to correctly prove all 15 equivalent pairs from the modules with beam width 1 and wider. With a beam width of 10, the WholeProof10 model proved 12. An example problem solvable by AxiomStep10 but not WholeProof10 is: $c(1A +B) = cB + cA$ which can be proven by applying the rewrite rules NeutralOp, DistributeRight, and Commute to the proper nodes. The WholeProof10 model mostly fails because it was not trained on how to apply repeated transformations at the same point in the AST. This suggests AxiomStep10 has generalized well to these hand-written problems.

\section{Related Work}
\label{sec:related}
%% \section{Related Work}
%% \label{sec:related}
%% \input{related}

%% \paragraph*{Program analysis with machine learning}

%% The work presented here is built upon two active research fields: program analysis
%% and generation using machine learning, and graph neural networks for analysis of
%% data which is best represented with graphs. There are recent survey papers
%% summarizing the current state of both of these fields \cite{AllamanisACM18} \cite{Wu19}.

\paragraph*{Theorem provers} The problem of equivalence as we formulated may be solved by other (smart) brute-force approaches, where a problem is solved by pathfinding. This ranges from theoreom proving systems like Coq \cite{bertot2013interactive} which supports the formal framework for equivalence we describe in this paper, to (Approximate Probabilistic) Model Checking \cite{clarke1994model,burch1992symbolic,herault2004approximate}, where a program equivalence system can also be built, e.g. \cite{steffen1991data,clarke2003behavioral,visser2003model,namjoshi2000syntactic}. Our contribution is not in the formal definition of program equivalence we presented, semantics-preserving rewrite systems have been studied, e.g. \cite{visser2004program,lucanu2015program,reddy1989rewriting}. But understanding why this particular formalism was well suited to deep learning graph-to-sequence systems was key.
The merits of stochastic search to accelerate such systems has been demonstrated, e.g. \cite{murawski2005probabilistic,herault2004approximate,gogate2012probabilistic}. The novelty of our approach is to develop carefully crafted graph-to-sequence neural networks to automatically learn an efficient pathfinding heuristic for this problem. Our approach is potentially applicable in these areas too, however training scalability can become a challenge if increasing the input representation size excessively. 
Theorem provers using deep learning have recently started to be investigated, Aygun et al.~\cite{Aygun20} developed a graph neural network system for automatic proof generation. Wu et al. \cite{Wu20} explores the ability of theorem provers using GNNs, TreeLSTMs, and BagOfWords architectures to generalize and solve proofs with lengths up to 7 axioms and found that GNNs performed the best of the architectures studied when more complex proofs were required. While our model works in a slightly different problem space, we study the ability of our models to generalize on proofs with lengths up to 10, with 14 different rewrite rules acting on 147 distinct axioms. These frameworks could also be used to prove equivalence between symbolic expressions, as theorem provers.

%% The paper "Learning to prove from synthetic theorems" \cite{Aygun20} created a form of graph neural networking with node, edge, and graph embeddings using a limited number of edge and node type, which contrasts with our 18 edge types that can allow for learning more problem-specific information at a moderate cost in added parameters. The show they can use synthetically generated theorems (not unlike our generator) to train a model which works well on human problems. Their figure 1 shows that the NN approaches (MLP as well as GNN) can prove more theorems than a basic theorem prover with a very simple cost function not based on neural networks. The basic model could generate more simple proofs quickly, but got passed over time by the neural network models that could do more complex proofs.

%% The paper " INT: An Inequality Benchmark for Evaluating Generalization in Theorem Proving" \cite{Wu20} looks at different models for performing proofs that are similar to ours. They show K/L parameters which I think line up with our 'number of axiom groups' of 14 and our 'maximum proof length' of 10. For proofs of length 10, we showed 66% success with 14 axiom groups; their tested models fall to 61% with K5 L7.

%% The paper " Towards Finding Longer Proofs" \cite{Zombori19} uses G to study Robinson Arithmetic, which tends to have long proofs but only has 6 axioms. (This arxiv paper isn't the same as the AITP 2 page summary you sent, but that didn't seem to be archival)

\paragraph*{Static program equivalence} Algorithms for static program equivalence have been developed, e.g. \cite{verdoolaege2012equivalence,alias2004recognition,barthou2002,iooss2014program}. These approaches typically restrict to demonstrating the equivalence of different schedules of the operations, possibly dynamically \cite{bao2016polycheck}. In this work we target graph-modifying rewrites (and therefore which alter the operation count). Barthou et 
al. \cite{barthou2002,alias2004recognition} have developed
techniques to recognize algorithm templates in programs. These
approaches are restricted to static/affine transformed programs.
Karfa et al. also designed a method that works for a subset of affine
programs using array data dependence graphs (ADDGs) to represent input
and transforming behaviors.  Operator-level equivalence checking
provides the capability to normalize expressions and establish
matching relations under algebraic transformations
\cite{karfa2013verification}.  Mansky and Gunter used the TRANS
language \cite{kalvala2009program} to represent transformations.  The
correctness proof implemented in the verification framework
\cite{mansky2010framework} is verified by the Isabelle
\cite{isabelle-web} proof assistant.  Other works
also include translation
validation \cite{sorin:proving,necula2000translation}.

\paragraph*{Program analysis with machine learning} Numerous prior work has employed (deep) machine learning for program analysis, e.g. \cite{AllamanisACM18,Alon19,Tufano19,LacomisDIRE2019,Raychev2015,Bavishi17}.
%% PHOG \cite{bielik16} presents a probabilistic
%% grammar to predict node types in an AST.
%
%% While this work does not use machine
%% learning on code, it presents some of the benefits of analyzing code with
%% probability distributions, which is the main strength that machine learning
%% is able to bring to a problem.
code2vec \cite{Alon19} teaches a method for
creating a useful embedding vector that summarizes the semantic
meaning of a snippet of code. Program repair approaches, e.g. \cite{Tufano19,Chen19} are deployed to automatically repair bugs in a program. Output accuracies of up to 20\% on the test set is reported, using sequence-to-sequence models.
Wang et al. \cite{Wang18} learns to extract the rules for Tomita grammars \cite{tomita82} with recurrent neural networks. The learned network weights are processed to create a verifiable deterministic finite automata (DFA) representation of the learned grammar. This work demonstrates that deterministic grammars can be learned with RNNs, which we rely on.

\paragraph*{Graph Neural Networks}
%% The organized nature of computer programs are well represented and analyzed
%% using graphs of various types - abstract syntax trees (ASTs), Control flow
%% graphs (CFGs), etc. \FIXME{Louis-Noel, do you want to cite a paper on that
%% statement?}.
Graph neural networks \cite{Scarselli09,Wu19} use machine learning
to analyze a set of nodes and edges for patterns related to a target problem.
Using a graph-to-sequence network with attention has been analyzed for natural
language processing \cite{Beck18}. Allamanis et al. use graph
neural networks to analyze code sequences and add edge types representing
LastUse, ComputedFrom, and LastWrite to improve the system's ability to
reason about the code \cite{AllamanisICLR18}. Their work achieves 
84\% accuracy on correcting variable misuse cases and provides insights
to useful edge types. %which may be valuable in extending pe-graph2seq to more complex programs.
Structure2vec \cite{Xu17} uses a graph neural network to detect binary code similarity. Structure2vec uses a graph
neural network to learn an embedding from a annotated control flow graph (ACFG)
of a program. This learning process targets the embedding so that equivalent
programs will have equivalent embeddings, reporting precision
scores of 84\% and 85\% on various test datasets for correctly predicting
program equivalence. It only outputs a probability of equivalence, and not a verifiable proof, which is sufficient in their context.
%% which supports our contention
%% that robust program equivalence work with machine learning must involve the
%% creation of an output that creates a proof sequence that confirms 2 programs
%% are equivalent.

%% In our work, each node, including the ProgramCompare node, learns an embedding
%% related to the task of generating a transformation sequence.
The G2SKGE model \cite{Li19} has a similar graph network structure which uses a node embedding (which they refer to as an information fusion mechanism) in order to predict
relationships between nodes. This technique of using a neural network to understand and predict
node interelationships is common to our approach.

%% There is also recent interest in reinforcement learning for proof generation \citep{Alhussein19,Bansal19}.

\section{Conclusion}
\label{sec:conclusion}
%% \section{Conclusion}
%% \label{sec:conclusion}
%% \input{conclusion}

In this work, we presented \texttt{pe-graph2axiom}, the first graph-to-sequence neural network system to generate verifiable axiomatic proofs (via rewrite rules) for equivalence for a class of symbolic programs. Evaluated on a rich language for linear algebra expressions, this system produces correct proofs of up to 10 axioms in length in 93\% of the 10,000 equivalent cases evaluated. We believe the performance of our approach comes in part from using graph neural networks for what they aim to excel at: learning efficient heuristics to quickly find paths in a graph; and the observation that program equivalence can be cast as a path-based solution that is efficiently found by such networks.

%% In this work, we presented \texttt{pe-graph2seq}, the first graph-to-sequence neural network system generating quickly verifiable program equivalence proofs. Evaluated on a rich language for linear algebra expressions, our system  outputs proofs when input programs are equivalent which are verified correct in 96\% of cases. In addition, the system always outputs non-equivalence for non-equivalent programs by design.

%% We believe the performance of our approach comes in part from using graph neural networks for what they aim to excel at: learning efficient heuristics to quickly find paths in a graph; and the observation that program equivalence can be cast as a path-based solution that is efficiently found by such networks. We demonstrated our approach on a carefully crafted linear algebra language, to expose clearly the various difficulties the system overcame, such as node deletion or subtree manipulation. We believe this has laid the foundations on how to build such deep learning systems for program equivalence in other languages.

%% %% by carefully co-designing the language axioms of equivalence and the example program generation tool. As long as the programs to be tested for equivalence can be modeled as an AST/graph structure, and that the axioms for equivalence can be expressed along our formalism, our approach would be applicable.

%% %% Acknowledgments
\begin{acks}                            %% acks environment is optional
This work was supported in part by the U.S. National Science Foundation award CCF-1750399.
%%                                         %% contents suppressed with 'anonymous'
%%   %% Commands \grantsponsor{<sponsorID>}{<name>}{<url>} and
%%   %% \grantnum[<url>]{<sponsorID>}{<number>} should be used to
%%   %% acknowledge financial support and will be used by metadata
%%   %% extraction tools.
%%   This material is based upon work supported by the
%%   \grantsponsor{GS100000001}{National Science
%%     Foundation}{http://dx.doi.org/10.13039/100000001} under Grant
%%   No.~\grantnum{GS100000001}{nnnnnnn} and Grant
%%   No.~\grantnum{GS100000001}{mmmmmmm}.  Any opinions, findings, and
%%   conclusions or recommendations expressed in this material are those
%%   of the author and do not necessarily reflect the views of the
%%   National Science Foundation.
\end{acks}

\balance
%% Bibliography
%\bibliography{lnp,pldi20,bibs/refs,bibs/gabriel,bibs/ierefs,bibs/ics15,iclr/paper/ICLR21.bib}
\bibliography{cc21}
%%% LNP: commented out bibs/lnp, i think it's duplicate with lnp.bib.

%% Appendix
\newpage

\appendix
\section{Appendix}
%% \FIXME{LNP: put your text-as is here. I will summarize, this is not essential for the paper. But it's great as appendix. Before submitting, we will need to harmonize notations between this section and the main paper. Ultra low priority (eg, friday stuff)}

%\FIXME{LNP: Absolutely IMPORTANT: we need to structure the appendix now, in sections at least, to enable correct ref in the main document}
{\centering\large{\textbf{Supplementary Materials:}\\Learning Axioms to Compute Verifiable\\ Symbolic Expression Equivalence Proofs\\ using Graph-to-Sequence Networks\\~}}
\vspace{-.5cm}
%%%%%%%%%
\section*{Document Overview}
\label{sec:appendix:intro}

This document supplements the submission \emph{Proving Equivalence Between Complex Expressions Using Graph-to-Sequence Neural Models}. We have provided below numerous additional information for completeness. We also provide access to anonymized software artifacts to replicate our results. Our supplementary materials are organized as follows:
\begin{itemize}[noitemsep,topsep=0pt,wide=0pt]
%\item Appendix~\ref{sec:appendix:motivation} of this document revisits the motivation and overview of our system in more depth.
%\item Appendix~\ref{sec:appendix:progequivframework} of this document formalizes the complete framework for axiomatic program equivalence we leverage in our work.
\item Appendix~\ref{sec:appendix:datasetgen} of this document presents the dataset generation approach we developed.
\item Appendix~\ref{sec:Axioms} of this document presents exhaustively the language for complex linear algebra expressions we evalaute on, including the list of all 147 axioms of equivalence we learned.
\item Appendix~\ref{sec:suppl:neuralnetwork} of this document presents additional details about the neural network architectures we developed.
\item The anonymized url \url{ https://gofile.io/d/IvqAnp} contains all trained models evaluated in this paper, including scripts to train them directly from our datasets, using OpenNMT. It contains also our code for generating datasets and training models, the testsets for our 4 key datasets, and key results files from our testset evaluations. 
\item Appendix~\ref{sec:suppl:additionalresults} of this document presents complementary experimental results and additional in-depth details on results presented in the main paper body.
\end{itemize}

%% \section{Motivation, Intuitions and System Overview}
%% \label{sec:appendix:motivation}

%% \input{appendix-motivation}

%% \section{Formal Framework for Program Equivalence}
%% \label{sec:appendix:progequivframework}

%% \input{appendix-progequivframework}

%\input{appendix-secs1-4}
%%%%%%%%%

\section{Dataset generation}
\label{sec:appendix:datasetgen}

%\section{Exp results}

%\section{Discussions}

\subsection{Generation of Examples}
\label{sec:appendix:genexamples}
% \FIXME{LNP: budget: max 1 page, target 0.75 or so. Note: to avoid conflicts on SVN you should edit this text here, but it will then be moved as last subsec of the previous section. Don't hesitate to create a separate latex file for it btw}

% Comes here the explanation of how you generated all programs and their transformations. Use a similar style as I suggest below using latex paragraphs to discuss the various elements, after having presented the overview/flow of steps at a high-level in 1-2 paragraphs (which themselves could be the description of a high-level algorithm or figure). The writing style is always ``problem was x, we implemented solution y, the motivating choice for this was z'' but in nice-flowing scientific English. Never hesitate to use math formulas and greek symbols to explain/summarize something, or pseudocode/algorithms, it's perfect for pldi.

%% As previously discussed, we identified program equivalence for linear algebra
%% as a valuable subset of the program equivalence problem.
Machine learning
benefits from large training sets, so in order to produce this data, we
created algorithms that would generate programs meeting a given language
grammar along with target programs which could be reached by applying a
given axiom set. By creating this process, we could create as large and
varied a dataset as our machine learning approach required. 

Algorithm \ref{alg:GenP1} provides an overview of the full program generation
algorithm. For this generation process, we define a set of operations and
operands on scalars, matrices, and vectors. For our process, we presume matrix
and vector dimensions are appropriate for the given operation as such dimension
checks are simple to implement and are not considered in our procedure. 
%%% LNP: Added a few more words, to reconcile different notations in the document.
Note the token syntax here is \emph{exactly} the one used by our system, 
%%% Added this:
and is \emph{strictly} semantically equivalent to the mathematical notations used to describe these operations, e.g. $1_{\mathbb{N}}$ is \texttt{1}.

\begin{itemize}[noitemsep,topsep=0pt,wide=0pt]
  \item Scalar operations: \texttt{+s -s *s /s is ns}, where \texttt{is} the unary reciprical and \texttt{ns} is the unary negation.
  \item Matrix operations: \texttt{+m -m *m im nm tm}, where \texttt{im} is matrix inversion, \texttt{nm} negates the matrix, and \texttt{tm} is matrix transpose.
  \item Vector operations: \texttt{+v -v *s nv}, where \texttt{nv} is the unary negation.
  \item Scalars: \texttt{a b c d e 0 1}
  \item Matrices: \texttt{A B C D E O I}, where \texttt{O} is the empty matrix and \texttt{I} is the identity matrix.
  \item Vectors: \texttt{v w x y z o}, where \texttt{o} is the empty vector.
  \item Summary: 16 operations, 20 terminal symbols
\end{itemize}

Initially, \texttt{GenP1} is called with \texttt{GenP1("+s -s *s /s +s -s *s /s +s -s *s /s is ns +m -m *m +m -m *m +m -m *m im nm tm +v -v *v +v -v *v +v -v *v nv",0.94)}"
  In this initial call binary operations are repeated so
that they are more likely to be created than unary operations, and the
initial probability that a child of the created graph node will itself
be an operation (as opposed to a terminal symbol) is set to
94\%. Since the algorithm subtracts a 19\% probability for children at
each level of the graph, trees are limited to 7 levels.

Algorithm \ref{alg:GenP1} starts execution by randomly selecting an
operation from the set provided as input. When \texttt{GenP1} is called
recursively, the operation set is limited such that the operation produces
the correct type as output (scalar, matrix, or vector). Lines 3 through 15
of the algorithm show an example case where the \texttt{*s} operation is
processed. This operation requires scalar operands. If the probability of
children at this level is met, then \texttt{GenP1} is called recursively
with only scalar operands available, otherwise a random scalar operand is
chosen. 

The text for algorithm \ref{alg:GenP1} does not show the process for all 
operations. Certain operations, such as \texttt{*v}, have a variety of 
operand types that can be chosen. The \texttt{*v} operand is a multiplication
which produces a vector. As such, $Av$ (matrix times vector), $bv$ (scalar
times vector), or $vc$ (vector times scalar) are all valid 
options and will be chosen randomly.

\begin{algorithm}[h!tb]
\DontPrintSemicolon
\SetAlgoLined
\KwResult{Prefix notation of computation with parenthesis}
\SetKwInOut{Input}{Input}\SetKwInOut{Output}{Output}
\Input{Ops, P}
\Output{(op L R) or (op L)}
\BlankLine
 
op = select randomly from Ops

// Create subtree for chosen op

\If{op == "*s"}{
    \eIf{random < P}{
        L = GenP1("+s -s *s /s +s -s *s /s is ns",P-0.19)
    }{
        L = select random scalar operand
    }
    \eIf{random < P}{
        R = GenP1("+s -s *s /s +s -s *s /s is ns",P-0.19)
    }{
        R = select random scalar operand
    }
    return (op L R)
}
// Other ops may have more complex options for children types.

// (For example, "*m" may have a matrix multiplied by a scalar or matrix)

...
% \If{op == "+v" or op == "-v"}{
%     \If{random < 0.5 and P > 0.1 and P < 0.5}{
%         // Bias creation to favor factorization
% 
%         L = GenP1("*v",P-0.23)
% 
%         R = GenP1("*v",P-0.23)
% 
%         return (op L R)
%     }
%     \eIf{random < P}{
%         L = GenP1("+v -v *v +v -v *v nv",P-0.23)
%     }{
%         L = select random vector operand
%     }
%     \eIf{random < P}{
%         R = GenP1("+v -v *v +v -v *v nv",P-0.23)
%     }{
%         R = select random vector operand
%     }
%     return (op L R)
% }
 
\caption{GenP1}
\label{alg:GenP1}
\end{algorithm}

After generating a program which follows the grammar rules of our language,
algorithm \ref{alg:GenP2} will produce a new program along with a set of
rewrite rules which transform the source program to the target program.

Algorithm \ref{alg:GenP2} receives as input the source program (or
subprogram) along with the \texttt{path} to the current root node of the
source program. If the source program is a terminal symbol, the algorithm
returns with no action taken. Otherwise, the program starts with an
operation and the algorithm proceeds to process options for transforming
the given operation. For our wholeproof10 and wholeproof5 datasets, algorithm \ref{alg:GenP2} is only called once, simplifying the possible node order and proof complexity. for the axiomstep10 and axiomstep5 datasets, algorithm \ref{alg:GenP2} is called multiple times, allowing for the possibility that after a path is chosen for one axiom any node can be accessed for the next axiom (including the same node).

As shown on line 10 of the algorithm, when the operation and children meet the
conditions necessary for a rewrite rule (in this case \texttt{NeutralOp}), the rule
is applied with some probability (in this case 50\%). Note that before 
processing a node, the left and right operands are further analyzed to 
determine their operators and operands as well (or $\bot$ if the child is a
terminal). Processing the left and right operands allows for complex axioms
to be applied, such as distribution or factorization. When a rule is
applied, the rewrite rule is added to the
rewrite rule sequence and 
a new target program is generated for any remaining subtrees. 
When creating the rewrite rules for subtrees, the \texttt{path} variable is updated as rewrites are done. In the case of \texttt{NeutralOp}, the current node is being updated, so the path is not changed. But in the case of the Commute rule, the return would be generated with \texttt{(op GenP2(R,path."left ") GenP2(L,path."right "))} which creates rewrite rules for the prior right and left operands of the \texttt{op} and updates the path used to the new node positions.
In order to analyze nearly equal programs, illegal rewrites can be optionally enabled;
for example, commuting a subtraction operation or mutating one operation into another.
In that case, the \texttt{GenP2} process continues to create a target program, but
\texttt{transform\_sequence} is set to \texttt{Not\_equal}.

%% \begin{figure}[h!tb]
\begin{algorithm}[h!tb]
\DontPrintSemicolon
\SetAlgoLined
\KwResult{Second program and transform\_sequence}
\SetKwInOut{Input}{Input}\SetKwInOut{Output}{Output}
\Input{P1, path}
\Output{P2}
\BlankLine

\If{terminal symbol}{return P1}

op = find operator of P1

L = find left operand of P1

R = find right operand of P1

Lop,LL,LR = operator and operands of left child

Rop,RL,RR = operator and operands of right child
 
// Randomly apply transform if allowed

\If{random < 0.5 and ((op == "+v" and (L == "o" or R == "o")) or (op == "-v" and R == "o"))}{
    append path."NeutralOp " to transform\_sequence

    // Eliminate unnecessary operator and 0 vector 

    \eIf{L == "o"}{
        return GenP2(R,path)
    }{
        return GenP2(L,path)
    }
}
 
\caption{GenP2}
\label{alg:GenP2}
\end{algorithm}

After these generation algorithms are run, a final data preparation
process is done which prunes the data set for the learning
algorithm. The pruning used on our final data set insures that the
$(P1,P2)$ program pair total to 100 tokens or fewer (where a
token is an operation or terminal), that the graph is such that every node is reachable from the root with a path of length 6 or less,
% 6 layers or fewer deep => dont talk about ast, we talk about graphs everywhere. adapted the def accordingly.
and that there are 10 or fewer rewrite rules applied. But within these restrictions, we assert that our random production rule procedure has a non-zero probability of producing any program allowed by the grammar. Also, the pruning insures that there are no
lexically equivalent programs in the process and removes some of the cases with fewer than 10
rewrite rules generated to bias the dataset to longer rewrite
sequences. Table \ref{tab:TransformPct} details the distribution of
rewrite rules created by the full process. Section \ref{sec:Axioms} details
all axioms when variable types and operators are considered.
%% Listings \ref{lst:equal}
%% and \ref{lst:notequal} show examples of 2 program pairs produced by
%% this process.

%% %% \noindent
%% \begin{figure}[h!tb]
%% \begin{minipage}{\linewidth}
%% \begin{lstlisting}[columns=flexible, frame=single, basicstyle=\footnotesize,label={lst:equal},caption={Example of equivalent program generation. In this case, the * operand is distributed over the - operand, and then "d * c" is commuted to "c * d".},captionpos=b,breaklines=true]
%% P1: ( d * ( ( ( - e ) / a ) - c ) ) 
%% P2: ( ( d * ( ( - e ) / a ) ) - ( c * d ) ) 
%% Rewrite rule sequence: DistributeRight right Commute
%% \end{lstlisting}
%% \end{minipage}
%% \end{figure}

%% %% \noindent
%% \begin{figure}[h!tb]
%% \begin{minipage}{\linewidth}
%% \begin{lstlisting}[columns=flexible, frame=single, basicstyle=\footnotesize,label={lst:notequal},caption={Example of non-equivalent generation. In this case, the first operation is changed and also a non-equivalent commutation of the "-" operation was done.},captionpos=b,breaklines=true]
%% P1: ( w - ( ( c * x ) - y ) ) 
%% P2: ( w + ( y - ( c * x ) ) ) 
%% Rewrite rule sequence: Not_equal
%% \end{lstlisting}
%% \end{minipage}
%% \end{figure}

We produce equivalent program samples by pseudo-randomly applying axioms on one randomly generated program to produce a rewrite sequence and the associated equivalent program. Given a randomly selected node in the program graph, our process checks which axiom(s) can be applied. E.g., the $+_m$ operator can have the Commute axiom applied, or depending on subtrees it may be allowed to have the Factorleft axiom applied, as discussed in Sec.~\ref{sec:expresults}. Generally we choose to apply or not an operator with 50\% probability, so that \texttt{pe-graph2axiom} is forced to rely on analysis of the two programs to determine whether an operator is applied instead of learning a bias due to the local node features. 

\subsection{Intermediate program generation}
The intermediate program generation algorithm is very similar to algorithm \ref{alg:GenP2}. For
program generation of the target program, algorithm \ref{alg:GenP2} will check that a node can
legally apply a given rule, apply the rule with some probability, record the action,
and process the remaining program. For intermediate program generation, we begin with a
$P1$ and a rewrite rule. We follow the path provided to identify the node,
check that a node can legally accept a rule, apply the
rule, and return the adjusted program.
If a rule cannot legally be applied, $P1$ is not successfully transformed.
If a rule can be legally applied to $P1$, the program is compared
lexically to $P2$ and if they match then equivalence has been proven.

\subsection{Complexity of Proving Equivalence}
\label{subsec:complexityofprovingequiv}
Table ~\ref{tab:count} shows the complexity of the solution space for our problem for proofs from our AxiomStep10 test dataset up to length 7 (deterministically computing all possible programs requires too many resources for longer proof lengths). The 'All possible nodes and axioms' row includes the total number of proofs of a given length available to our problem space. The entry 5933 for a single axiom represents that for an AST depth of 7 we have 43 axioms which can be applied to all 63 possible operator nodes and 104 axioms which can be applied to the 31 nodes which possibly have child operator nodes themselves: 63*43+31*104=5933. Subsequent columns can select repeatedly from the same set growing as $5933^2$ to $5933^7$. The 'sample node + axiom group' row is based on our 10,000 sample test dataset and represents the possible selection of any of the 14 axiom groups being applied to any node in the program. The 'sample node + legal axiom' row represents only legal node plus legal axiom group being applied and effectively represents the total number of programs derivable from the start program in the test dataset. The final row 'Sample derivable unique programs' represents the total number of programs derived from legal node and axiom sequences which are lexically unique.

%\vspace{-.1cm}
\begin{table*}[h!tb]
\vspace{-.5cm}
  \caption{Counts for equivalence proof possibilities\vspace{-.3cm}}
  \label{tab:count}
  \small
  \centering
  \begin{tabular}{@{}lrrrrrrrr@{}}
    \toprule
    & & \multicolumn{7}{c}{Proof length in axioms} \\
    Proof description & & 1 & 2 & 3 & 4 & 5 & 6 & 7  \\
    \cmidrule{1-1} \cmidrule{3-9}
    All Possible nodes and axioms & & 5933 & 3.5E+07 & 2.1E+11 & 1.2E+15 & 7.4E+18 & 4.4E+22 & 2.6E+26 \\
    Sample Node + Any Axiom  & & 226 & 46900 & 1.5E+07 & 8.8E+09 & 5.0E+12 & 3.3E+15 & 2.7E+18 \\
	Sample Node + Legal Axiom  & & 11.2 & 77.8 & 931 & 15812 & 3.4E+05 & 8.2E+06 & 1.8E+08 \\
	Unique Programs from Sample & & 9.2 & 47.4 & 264 & 1574 & 10052 & 65176 & 4.6E+05 \\
    \bottomrule
  \end{tabular}
\end{table*}

\section{Language and Axioms for Complex Linear Algebra Expressions}
\label{sec:Axioms}
%\section{Full axiom list -- to be renamed, full LA language description instead of just axioms. Do NOT change the sec. label!}
We now provide the complete description of the input language for multi-type linear algebra expressions we use to evaluate our work, and the complete list of all axioms that are used to compute equivalence between programs.

\paragraph*{Variable types} We model programs made of scalars, vectors and matrices. We limit programs to contain no more than 5 distinct variable names of each type in a program:
\begin{itemize}[noitemsep,topsep=0pt,wide=0pt]
\item Scalar variables are noted $a$, $b$, ..., $e$.
\item Vector variables are noted $\vec v$, $\vec w$, ..., $\vec z$.
\item Matrix variables are noted $A$, $B$, ..., $E$.
\end{itemize}

Note we also explicitly distinguish the neutral and absorbing elements for scalars and matrices, e.g. \texttt{1} $= 1_{\mathbb{N}}$. This enables the creation of simplification of expressions as a program equivalence problem, e.g. if $A+B-(B+A)=0_{\mathbb{K}\times\mathbb{K}}$

\paragraph*{Unary operators} We model 6 distinct unary operators, all applicable to any variable of the appropriate type:
\begin{itemize}[noitemsep,topsep=0pt,wide=0pt]
\item \texttt{is(a)} $= a^{-1}$ is the unary reciprocal for scalars, \texttt{im(A)} $= A^{-1}$ is matrix inverse.
\item \texttt{ns(a)} $= -a$ is unary negation for scalars, \texttt{nv(v)} $= -\vec v$ for vectors, \texttt{nm(M)} $ = -M$ for matrices.
\item \texttt{tm(M)} $= M^t$ is matrix transposition.
\end{itemize}

\paragraph*{Binary operators} We model 10 distinct binary operators that operate on two values. 7 operators require the same type for both operands, while 3 enable multi-type operands (e.g., scaling a matrix by a scalar). Note we do not consider potential vector/matrix size compatibility criterion for these operators, in fact we do not represent vector or matrix sizes at all in our language, for simplicity.
\begin{itemize}[noitemsep,topsep=0pt,wide=0pt]
\item \texttt{+s(a, b)}$= a+b$, the addition on scalars, along with \texttt{-s(a,b)} $ = a-b$, \texttt{*s(a,b)} $ = a*b$ and \texttt{/s(a,b)} $ = a/b$.
\item \texttt{+v( v, w)} $ = \vec v + \vec w$, the addition on vectors, along with \texttt{-v( v , w)} $ = \vec v - \vec w$, \texttt{*v( v,  w)} $ = {\vec v} . {\vec w}$ the dot product between two vectors, producing a scalar.
\item \texttt{+m(A, B)} $ = A+B$, the addition on matrices, along with \texttt{-m(A, B)} $ = A-B$,  and \texttt{*m(A, B)} $ = AB$ the product of matrices. 
\item \texttt{*m(a,A)} $= a\dot A $ and \texttt{*m(A,a)} $=A\dot a$ are used to represent scaling a matrix by a scalar.
\item \texttt{*m(v,A)} $= \vec v A $ represents a vector-matrix product.
\item \texttt{*v(a,v)} $= a \vec v$ and \texttt{*v(v,a)} $= \vec v  a$ represent scaling a vector by a scalar.
\end{itemize}

\paragraph*{List of axioms of equivalence}

Tables~\ref{tab:TransformAll1}-\ref{tab:TransformAll2} show the full 147 axioms supported by our rewrite rules. Many rewrite rules can be applied to all 3 variable types as well as multiple operator types. 
%As those tables are very large, they are displayed
%% \FloatBarrier

{
\begin{table*}[h!tb]
{\small
\begin{minipage}[t]{\textwidth}
\begin{minipage}[t]{.49\textwidth}
%\begin{tabular}{ |p{2.4cm}|p{1cm}|p{3.4cm}| }
\begin{tabular}[t]{ l|l|l }

 \hline
 Rewrite Rule & ID & Example(s) \\
 \hline
 Cancel & 1 & (a - a) $\rightarrow$ 0  \\
        & 2 & (b/b) $\rightarrow$ 1  \\
        & 3 & (A - A) $\rightarrow$ O  \\
        & 4 & $(A * A^{-1}) \rightarrow$ I  \\
        & 5 & $(A^{-1} * A) \rightarrow$ I  \\
        & 6 & (v - v) $\rightarrow$ o  \\
 \hline
 NeutralOp  & 7  & (a + 0) $\rightarrow$ a  \\
       & 8  & (0 + a) $\rightarrow$ a  \\
       & 9  & (a - 0) $\rightarrow$ a  \\
       & 10 & (a * 1) $\rightarrow$ a  \\
       & 11 & (1 * a) $\rightarrow$ a  \\
       & 13 & (a / 1) $\rightarrow$ a  \\
       & 14 & (A + O) $\rightarrow$ A  \\
       & 15 & (O + A) $\rightarrow$ A  \\
       & 16 & (A - O) $\rightarrow$ A  \\
       & 17 & (A * I) $\rightarrow$ A  \\
       & 18 & (I * A) $\rightarrow$ A  \\
       & 19 & (v + o) $\rightarrow$ v  \\
       & 20 & (o + v) $\rightarrow$ v  \\
       & 21 & (v - o) $\rightarrow$ v  \\
 \hline
 DoubleOp & 22 & -(-a)) $\rightarrow$ a \\
        & 23 & $(a^{-1})^{-1} \rightarrow$ a \\
        & 24 & $-(-A) \rightarrow A$ \\
        & 25 & $(A^{-1})^{-1} \rightarrow A$ \\
        & 26 & $(A^{t})^{t} \rightarrow A$ \\
        & 27 & $-(-v)) \rightarrow v$ \\
 \hline
 DistributeRight & 64 & a(b + c) $\rightarrow$ ab + ac  \\
                 & 65 & a(b - c) $\rightarrow$ ab - ac  \\
                 & 66 & a(v + w) $\rightarrow$ av + av  \\
                 & 67 & a(v - w) $\rightarrow$ av - av  \\
                 & 68 & A(B + C) $\rightarrow$ AB + AC  \\
                 & 69 & A(B - C) $\rightarrow$ AB - AC  \\
                 & 70 & a(B + C) $\rightarrow$ aB + aC  \\
                 & 71 & a(B - C) $\rightarrow$ aB - aC  \\
 \hline
\end{tabular}
\end{minipage}
\begin{minipage}[t]{.49\textwidth}
\begin{tabular}[t]{ l|l|l }
\hline
 Rewrite Rule & ID & Example(s) \\
 \hline
 AbsorbOp & 28 & (a * 0) $\rightarrow$ 0  \\
          & 29 & (0 * a) $\rightarrow$ 0  \\
          & 30 & (A * 0) $\rightarrow$ O  \\
          & 31 & (0 * A) $\rightarrow$ O  \\
          & 32 & (A * O) $\rightarrow$ O  \\
          & 33 & (O * A) $\rightarrow$ O  \\
          & 34 & (A * o) $\rightarrow$ o  \\
          & 35 & (a * o) $\rightarrow$ o  \\
          & 36 & (o * a) $\rightarrow$ o  \\
          & 37 & (0 * v) $\rightarrow$ o  \\
          & 38 & (v * 0) $\rightarrow$ o  \\
          & 39 & (O * v) $\rightarrow$ o  \\ \hline
 Commute & 40 & (a + b) $\rightarrow$ (b + a)  \\
         & 41 & (a * b) $\rightarrow$ (b * a)  \\
         & 42 & (A + B) $\rightarrow$ (B + A)  \\
         & 43 & (A * a) $\rightarrow$ (a * A)  \\
         & 44 & (a * A) $\rightarrow$ (A * A)  \\
         & 45 & (A * O) $\rightarrow$ (O * A)  \\
         & 46 & (O * A) $\rightarrow$ (A * O)  \\
         & 47 & (A * I) $\rightarrow$ (I * A)  \\
         & 48 & (I * A) $\rightarrow$ (A * I)  \\
         & 49 & (v + w) $\rightarrow$ (w + v)  \\
         & 50 & (v * a) $\rightarrow$ (a * v)  \\
         & 51 & (a * v) $\rightarrow$ (v * a)  \\
 \hline
 DistributeLeft & 52 & (a + b)c $\rightarrow$ ac + bc  \\
                & 53 & (a - b)c $\rightarrow$ ac - bc  \\
                & 54 & (a + b)/c $\rightarrow$ a/c + b/c  \\
                & 55 & (a - b)/c $\rightarrow$ a/c - b/c  \\
                & 56 & (v + w)*a $\rightarrow$ va + wa  \\
                & 57 & (v - w)*a $\rightarrow$ va - wa  \\
                & 58 & (A + B)C $\rightarrow$ AC + BC  \\
                & 59 & (A - B)C $\rightarrow$ AC - BC  \\
                & 60 & (A + B)v $\rightarrow$ Av + Bv  \\
                & 61 & (A - B)v $\rightarrow$ Av - Bv  \\
                & 62 & (A + B)a $\rightarrow$ Aa + Ba  \\
                & 63 & (A - B)a $\rightarrow$ Aa - Ba  \\
 \hline
\end{tabular}
\end{minipage}
\end{minipage}

}
\caption{Full axiom count when all type options and other supported permutations are included (part 1 of 2)\label{tab:TransformAll1}}
\end{table*}
}

\begin{table*}[h!tb]
{\small
%\begin{tabular}{ |p{2.4cm}|p{1cm}|p{3.4cm}| }

\begin{minipage}[t]{\textwidth}
\begin{minipage}[t]{.49\textwidth}

\begin{tabular}[t]{ l|l|l }
 \hline
 Rewrite Rule & ID & Example(s) \\
 \hline
 FactorLeft & 72 & ab + ac $\rightarrow$ a(b+c) \\
            & 73 & ab - ac $\rightarrow$ a(b-c) \\
            & 74 & AB + AC $\rightarrow$ A(B+C) \\
            & 75 & AB - AC $\rightarrow$ A(B-C) \\
            & 76 & Av + Aw $\rightarrow$ A(v+w) \\
            & 77 & Av - Aw $\rightarrow$ A(v-w) \\
            & 78 & Aa + Ab $\rightarrow$ A(a+b) \\
            & 79 & Aa - Ab $\rightarrow$ A(a-b) \\
            & 80 & va + vb $\rightarrow$ v(a+b) \\
            & 81 & va - vb $\rightarrow$ v(a-b) \\
 \hline
 FactorRight & 82 & ac + bc $\rightarrow$ (a+b)c \\
             & 83 & ac - bc $\rightarrow$ (a-b)c \\
             & 84 & a/c + b/c $\rightarrow$ (a+b)/c \\
             & 85 & a/c - b/c $\rightarrow$ (a-b)/c \\
             & 86 & AC + BC $\rightarrow$ (A+B)C \\
             & 87 & AC - BC $\rightarrow$ (A-B)C \\
             & 88 & Av + Bv $\rightarrow$ (A+B)v \\
             & 89 & Av - Bv $\rightarrow$ (A-B)v \\
             & 90 & Aa + Ba $\rightarrow$ (A+B)a \\
             & 91 & Aa - Ba $\rightarrow$ (A-B)a \\
             & 92 & va + wa $\rightarrow$ (v+w)a \\
             & 93 & va - wa $\rightarrow$ (v-w)a \\
 \hline
 AssociativeLeft & 94 & a+(b+c) $\rightarrow$ (a+b)+c  \\
                 & 95 & a+(b-c) $\rightarrow$ (a+b)-c  \\
                 & 96 & a(bc) $\rightarrow$ (ab)c  \\
                 & 97 & a(b/c) $\rightarrow$ (ab)/c  \\
                 & 98 & A+(B+C) $\rightarrow$ (A+B)+C  \\
                 & 99 & A+(B-C) $\rightarrow$ (A+B)-C  \\
                 & 100 & A(BC) $\rightarrow$ (AB)C  \\
                 & 101 & A(Ba) $\rightarrow$ (AB)a  \\
                 & 102 & A(aB) $\rightarrow$ (Aa)B  \\
                 & 103 & a(AB) $\rightarrow$ (aA)B  \\
                 & 104 & A(Bv) $\rightarrow$ (AB)v  \\
                 & 105 & A(va) $\rightarrow$ (Av)a  \\
                 & 106 & A(av) $\rightarrow$ (Aa)v  \\
                 & 107 & a(Av) $\rightarrow$ (aA)v  \\
                 & 108 & v+(w+x) $\rightarrow$ (v+w)+x  \\
                 & 109 & v+(w-x) $\rightarrow$ (v+w)-x  \\
                 & 110 & v(ab) $\rightarrow$ (va)b  \\
                 & 111 & a(vb) $\rightarrow$ (av)b  \\
                 & 112 & a(bv) $\rightarrow$ (ab)v  \\
 \hline
\end{tabular}
\end{minipage}
\begin{minipage}[t]{.49\textwidth}
%\begin{tabular}{ |p{2.4cm}|p{1cm}|p{3.4cm}| }
\begin{tabular}[t]{ l|l|l }
 \hline
 Rewrite Rule & ID & Example(s) \\
 \hline
 AssociativeRight & 113 & (a+b)+c $\rightarrow$ a+(b+c)  \\
                  & 114 & (a+b)-c $\rightarrow$ a+(b-c)  \\
                  & 115 & (ab)c $\rightarrow$ a(bc)  \\
                  & 116 & (A+B)+C $\rightarrow$ A+(B+C)  \\
                  & 117 & (A+B)-C $\rightarrow$ A+(B-C)  \\
                  & 118 & (AB)C $\rightarrow$ A(BC)  \\
                  & 119 & (AB)a $\rightarrow$ A(Ba)  \\
                  & 120 & (Aa)B $\rightarrow$ A(aB)  \\
                  & 121 & (aA)B $\rightarrow$ a(AB)  \\
                  & 122 & (Av)a $\rightarrow$ A(va)  \\
                  & 123 & (Aa)v $\rightarrow$ A(av)  \\
                  & 124 & (aA)v $\rightarrow$ a(Av)  \\
                  & 125 & (va)b $\rightarrow$ v(ab)  \\
                  & 126 & (av)b $\rightarrow$ a(vb)  \\
                  & 127 & (ab)v $\rightarrow$ a(bv)  \\
                  & 128 & (v+w)+x $\rightarrow$ v+(w+x)  \\
                  & 129 & (v+w)-x $\rightarrow$ v+(w-x)  \\
 \hline
 FlipLeft & 130 & -(a - b) $\rightarrow$ b-a \\
          & 131 & $(a / b)^{-1} \rightarrow$ b/a \\
          & 132 & $-(A - B) \rightarrow$ (B - A) \\
          & 133 & $-(v - w) \rightarrow$ (w - v) \\
 \hline
 FlipRight & 134 & a/(b/c) $\rightarrow$ a(c/b)  \\
           & 135 & $a/(b^{-1}) \rightarrow$ ab  \\
           & 136 & a-(b-c) $\rightarrow$ a+(c-b)  \\
           & 137 & a-(-b) $\rightarrow$ a+b  \\
           & 138 & A-(B-C) $\rightarrow$ A+(C-B)  \\
           & 139 & A-(-B) $\rightarrow$ A+B  \\
           & 140 & v-(w-x) $\rightarrow$ v+(x-w)  \\
           & 141 & v-(-w) $\rightarrow$ v+w \\
 \hline
 Transpose & 142 & $(AB) \rightarrow (B^{t}A^{t})^t$  \\
           & 143 & $(A+B) \rightarrow (A^{t}+B^{t})^t$  \\
           & 144 & $(A-B) \rightarrow (A^{t}-B^{t})^t$  \\
           & 145 & $(AB)^{t} \rightarrow B^{t}A^{t}$  \\
           & 146 & $(A+B)^{t} \rightarrow A^{t}+B^{t}$  \\
           & 147 & $(A-B)^{t} \rightarrow A^{t}-B^{t}$  \\
 \hline
\end{tabular}

\end{minipage}
\end{minipage}

}
\caption{Full axiom count when all type options and other supported permutations are included (part 2 of 2)\label{tab:TransformAll2}}
\end{table*}

\section{Details on neural network model}
\label{sec:suppl:neuralnetwork}

\begin{figure*}
\includegraphics[width=14cm]{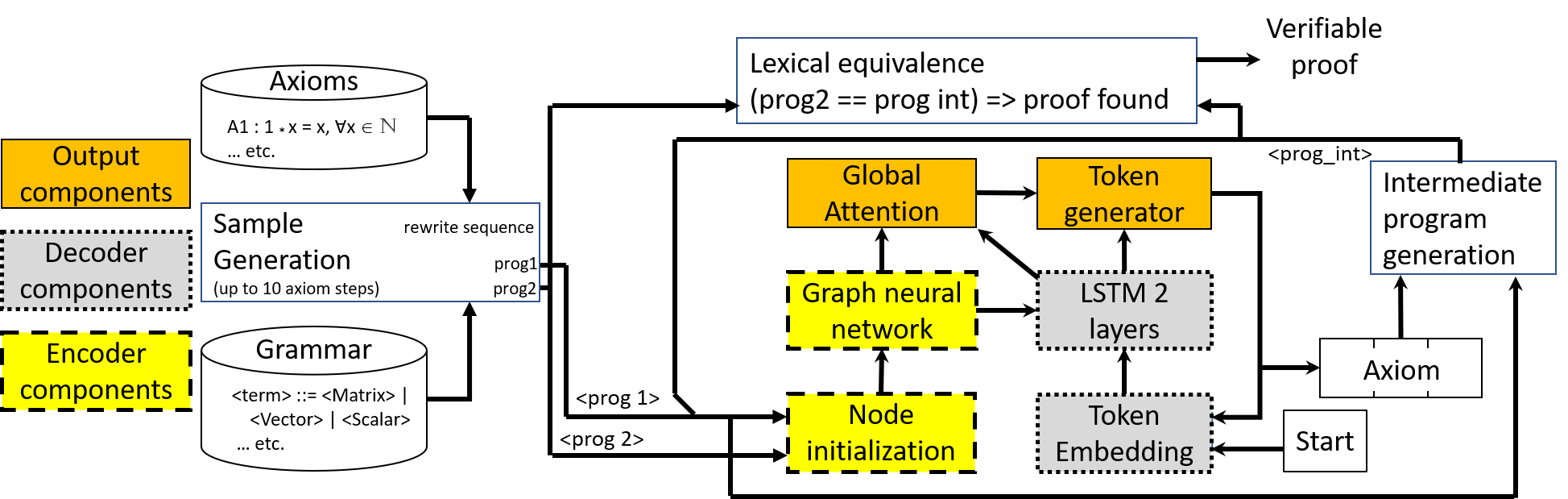}
\caption{\texttt{pe-graph2axiom} System Overview}
\label{fig:Full:appendix}
\end{figure*}

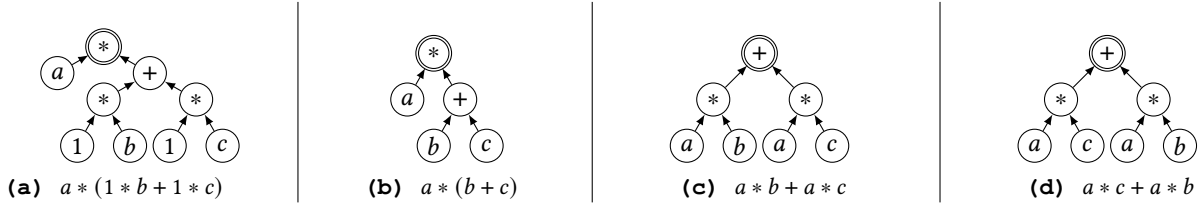
\begin{figure*}
    \centering

    %% \begin{subfigures}
    \centering
      \subfloat[][$a*(1* b+1* c)$]{
      \begin{minipage}[b]{0.25\textwidth}
        \label{fig:appendix:treeexamples:1}
        \centering
        \begin{tikzpicture} [level distance=1em, inner sep=1pt, minimum size=1.25em, edge from parent/.style={draw,latex-}]
            \node [circle, double, draw] {$*$}
            child [sibling distance=3.5em] {node [circle, draw] {$a$}
            child{edge from parent[draw=none] node [opacity=0] {}
            child  [sibling distance=2em]{edge from parent[draw=none] node [opacity=0] {}}
            child  [sibling distance=2em]{edge from parent[draw=none] node [opacity=0] {}}
            }
            child{edge from parent[draw=none] node [opacity=0] {}}
            }
            child [sibling distance=3.5em] {node [circle, draw] {$+$}
            child [sibling distance=3.5em, level distance=1em] {node [circle, draw] {$*$}
            child [sibling distance=2em, level distance=1.75em] {node [circle, draw] {$1$}}
            child [sibling distance=2em, level distance=1.75em] {node [circle, draw] {$b$}}
            }
            child [sibling distance=3.5em, level distance=1em] {node [circle, draw] {$*$}
            child [sibling distance=2em, level distance=1.75em] {node [circle, draw] {$1$}}
            child [sibling distance=2em, level distance=1.75em] {node [circle, draw] {$c$}}
            }
            };
        \end{tikzpicture}
        %\caption{$a*(1* b+1* c)$}        
        \end{minipage}
      %% \end{subfigures}
      }
      \unskip\ \hfill \vrule \hfill
      \subfloat[][$a*(b+c)$]{
    %% \begin{subfigures}
      \begin{minipage}[b]{0.20\textwidth}
        \centering
        \begin{tikzpicture} [level distance=1.75em, inner sep=1pt, minimum size=1.25em, sibling distance=2em, edge from parent/.style={draw,latex-}]
            \node [circle, double, draw, align=center] {$*$}
                child {node [circle, draw] {$a$}
                    child{edge from parent[draw=none] node [opacity=0] {}}
                    child{edge from parent[draw=none] node [opacity=0] {}}
                }
                child {node [circle, draw] {$+$}
                    child{node [circle, draw] {$b$}}
                    child {node [circle, draw] {$c$}}
                };
        \end{tikzpicture}
        %% \caption{$a*(b+c)$}
        \label{fig:appendix:treeexamples:2}
        \end{minipage}
      %% \end{subfigures}
      }
    \unskip\ \hfill \vrule \hfill
    %% \begin{subfigures}
    \subfloat[][$a * b + a * c$]{
      \begin{minipage}[b]{0.24\textwidth}
        \centering
        \begin{tikzpicture} [level distance=1.75em, inner sep=1pt, minimum size=1.25em, sibling distance=3.5em, edge from parent/.style={draw,latex-}]
            \node [circle, double, draw] {$+$}
                child {node [circle, draw] {$*$}
                    child [sibling distance=2em] {node [circle, draw] {$a$}}
                    child [sibling distance=2em] {node [circle, draw] {$b$}}
                }
                child {node [circle, draw] {$*$}
                    child [sibling distance=2em] {node [circle, draw] {$a$}}
                    child [sibling distance=2em] {node [circle, draw] {$c$}}
                };
        \end{tikzpicture}
        %% \caption{$a * b + a * c$}
        \label{fig:appendix:treeexamples:3}
        \end{minipage}
      %% \end{subfigures}
      }
    \unskip\ \hfill \vrule \hfill
    \subfloat[][$a * c + a * b$]{
    %% \begin{subfigures}
      \begin{minipage}[b]{0.24\textwidth}
        \centering
        \begin{tikzpicture} [level distance=1.75em, inner sep=1pt, minimum size=1.25em, sibling distance=3.5em, edge from parent/.style={draw,latex-}]
            \node [circle, double, draw] {$+$}
                child {node [circle, draw] {$*$}
                    child [sibling distance=2em] {node [circle, draw] {$a$}}
                    child [sibling distance=2em] {node [circle, draw] {$c$}}
                }
                child {node [circle, draw] {$*$}
                    child [sibling distance=2em] {node [circle, draw] {$a$}}
                    child [sibling distance=2em] {node [circle, draw] {$b$}}
                };
        \end{tikzpicture}
        %% \caption{$a * c + a * b$}
        \label{fig:appendix:treeexamples:4}
      \end{minipage}
      }
    \caption{Examples of Computations}
    \label{fig:appendix:treeexamples}
    %% \end{subfigures}

\end{figure*}

%%%%%%%%%%%%%%%%%%
%%%%%%%%%%%%%%%%%%

Figure \ref{fig:Full:appendix} overviews the entire \texttt{pe-graph2axiom} architecture including sample generation, the graph-to-sequence network, the intermediate program generation, and lexical equivalence checker. In this section we will discuss the implementation details of these components.

\paragraph*{Graph neural network internal representation}
The sample generation discussed in section \ref{sec:samplegen}
provides input to the \textsf{Node Initialization} module in
Fig.~\ref{fig:Full:appendix} to create the initial state of our graph neural
network. For each node in the program graph, a node will be initialized in our
graph neural network. Each node has a hidden state represented by a
vector of 256 floating point values which are used to create an
embedding for the full meaning of the given node. Initially all 256 
dimensions of the hidden states of the nodes are set to zero except for 2. 
Given $N$ tokens in our input program language, one of the dimensions from 
1 through $N$ of a node will be set based on the token at the program position 
that the node represents. For example, if the scalar variable $a$ is assigned to
be token 3 in our language, then the $a$ nodes of Fig.~\ref{fig:appendix:treeexamples} recalled below
would have their 3rd dimension initialized to 1.0. This is a one-hot encoding similar to that used
in neural machine translation models which leverage Word2vec \cite{Mikolov13}. The second non-zero
dimension in our node initialization indicates the tree depth, with the root for the program being at depth 1. We set the dimension $N$+$depth$ to 1.0; hence, the $a$ nodes in Fig~\ref{fig:appendix:treeexamples}, which vary from level 2 or 3 in the graph, would set dimension $N+2$ or $N+3$ to 1.
In addition to nodes correlating to all tokens in both input programs, we initialize
a root node for program comparison which has edges connecting to the root nodes of both programs.  The root node does not represent a token from the language, but it is initialized with a 1.0 in a hidden state dimension reserved for its identification.

For a graph neural network, the edge connections between nodes are a
crucial part of the setup. In particular, to match the formulation of our problem, we must ease the ability of the network to walk the input program graphs. We therefore designed a unified graph input, where both program graphs are unified in a single graph using a single connecting root node; and where additional edges are inserted to make the graph fully walkable.

\begin{figure*}[h!tb]
\vspace{-.5cm}
\centering
\includegraphics[width=0.8\textwidth]{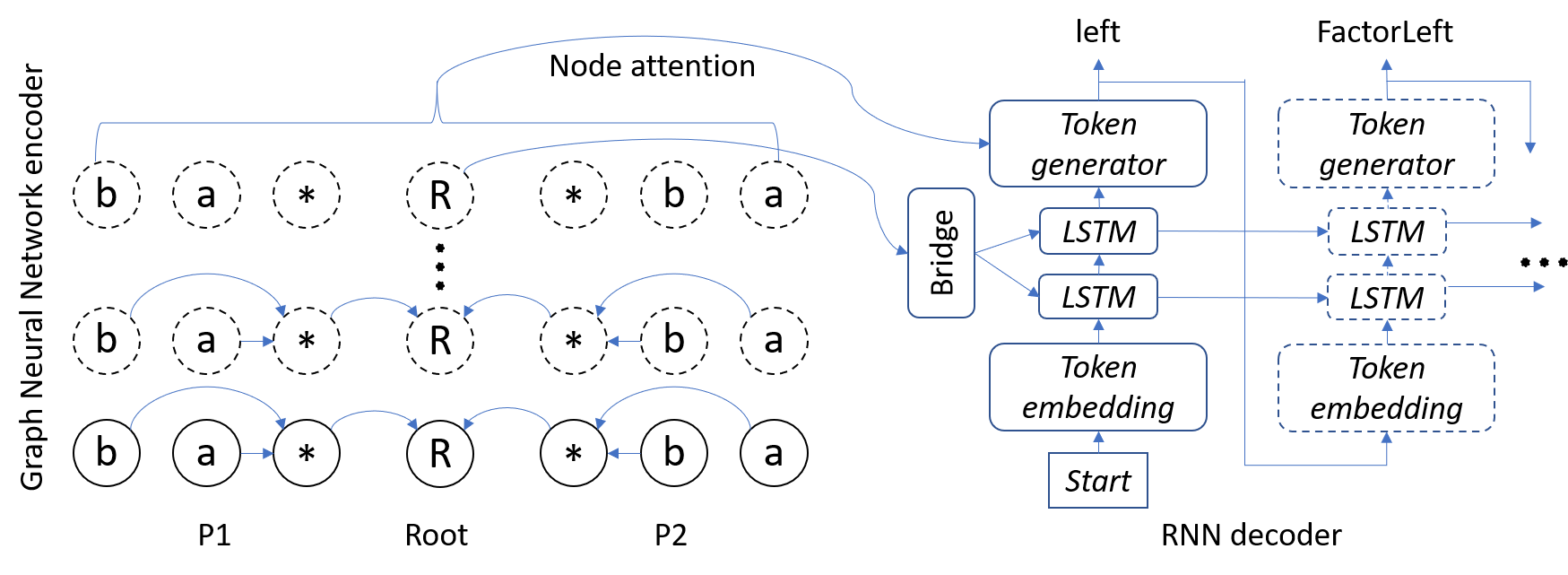}
\caption{Graph-to-sequence neural network data flow details.}
\label{fig:Network}
\vspace{-.5cm}
\end{figure*}

In our full model, we support 9 edge types and their reverse edges. The edge types are: 1) left child of binary
op, 2) right child of binary op, 3) child of unary op, 4) root node to
program 1, 5) root node to program 2, 6-9) there are 4 edge
types for the four node grandchilden (LL, LR, RL, RR). After the node
hidden states and edge adjacency matrix are initialized, the network is
ready to begin processing. This initial state is indicated in
figure \ref{fig:Network} by the solid circles in the lower left of the
diagram.

\paragraph*{Beam search}
A typical approach when using sequence-to-sequence systems is to
enable \emph{beam search}, the process of asking for multiple answers to
the same question to the network. It is particularly relevant when
creating outputs which can be automatically
checked \cite{Chen19,ahmed18}. Beam search can be viewed
 as proposing multiple possible axioms to apply. Given
the stochastic nature of generation model, a beam width of $n$ can
be thought of as creating the $n$ most likely sequences given
the training data the model as learned on. Each proposal
can be checked for validity, the first valid one is outputted by the
system, demonstrating equivalence. Our system builds on the neural network beam search provided by OpenNMT to create a 'system beam search' of variable width. In particular, we set the OpenNMT network beam search to 3, which constrains the token generator to produce 3 possible axiom/node proposals for a given pair of input programs. Using these 3 proposals, when our system beam width is 10, we build up to 10 intermediate programs that are being processed in the search for a proof. To illustrate with a system beam width of 5, after $P1$ and $P2$ are provided to the neural network, 3 possible intermediate programs may be created (so long as all axioms are legal and don't produce duplicates). After those 3 intermediates are processed, 9 possible new intermediates are created, all of which are checked for lexical equivalence with $P2$, but only 5 of which are fed back into the neural network for further axiom generation. This process is continued for up to 12 axioms at which point the system concludes an equivalence proof cannot be found and the programs are likely not equivalent.
We evaluate in
Sec.~\ref{sec:expresults} beam sizes ranging from 1 to 10, 
showing higher success with larger beams.

\section{Details on Experimental Results}
\label{sec:suppl:additionalresults}

\subsection{Complementary Results and Observations}
Table~\ref{tab:hyperparameter} describes part of our neural network hyperparameter tuning showing that our golden model has as high a result as other variations explored. Note that the validation token accuracy is not too high (it's not above 90\%) despite the ability to predict full correct proofs with over 93\% accuracy. This is because the training dataset can have multiple examples of axioms given similar input programs. For example, proving "(a+b)(c+d) = (b+a)(d+c)" requires commuting the left and right subexpressions. The training dataset could have similar programs which are sometimes transformed first with a right Commute and then a left or vice-versa. Given this data, the network would learn to apply one or the other (it would not get trained to use associativity for these program pairs for example), hence the actual output given may or may not match the validation target axiom. We will discuss this further in section~\ref{sec:appendix:multiaxiom}.

\begin{table}[h!tb]
  \caption{Hyperparameter experiments. Summary of best validation token accuracy result after 2 runs for up to 100,000 training iterations. The golden model has 256 graph nodes and decoder dimensions, 2 decoder LSTM layers, starts training with a learning rate of 0.8, and uses 10 steps to stabilize the GGNN encoder.}
  \label{tab:hyperparameter}
  \small
  \centering
  \begin{tabular}{@{}lrr@{}}
    \toprule
    Parameter & Value & Validation \\
    & & token accuracy \\
    \midrule
    Golden model & & 83.89 \\
    Graph node+decoder LSTM dimension & 192 & 83.89 \\
                                          & 320 & 83.58 \\
    Decoder LSTM layers & 1 & 83.53 \\
    Initial learning rate & 0.75 & 83.76 \\
                          & 0.85 & 83.57 \\
    GGNN stability steps  & 12 & 83.19 \\
                          & 8 & 83.61 \\
    \bottomrule
  \end{tabular}
\end{table}

\begin{figure}[h!tb]
\centering
\includegraphics[width=8cm]{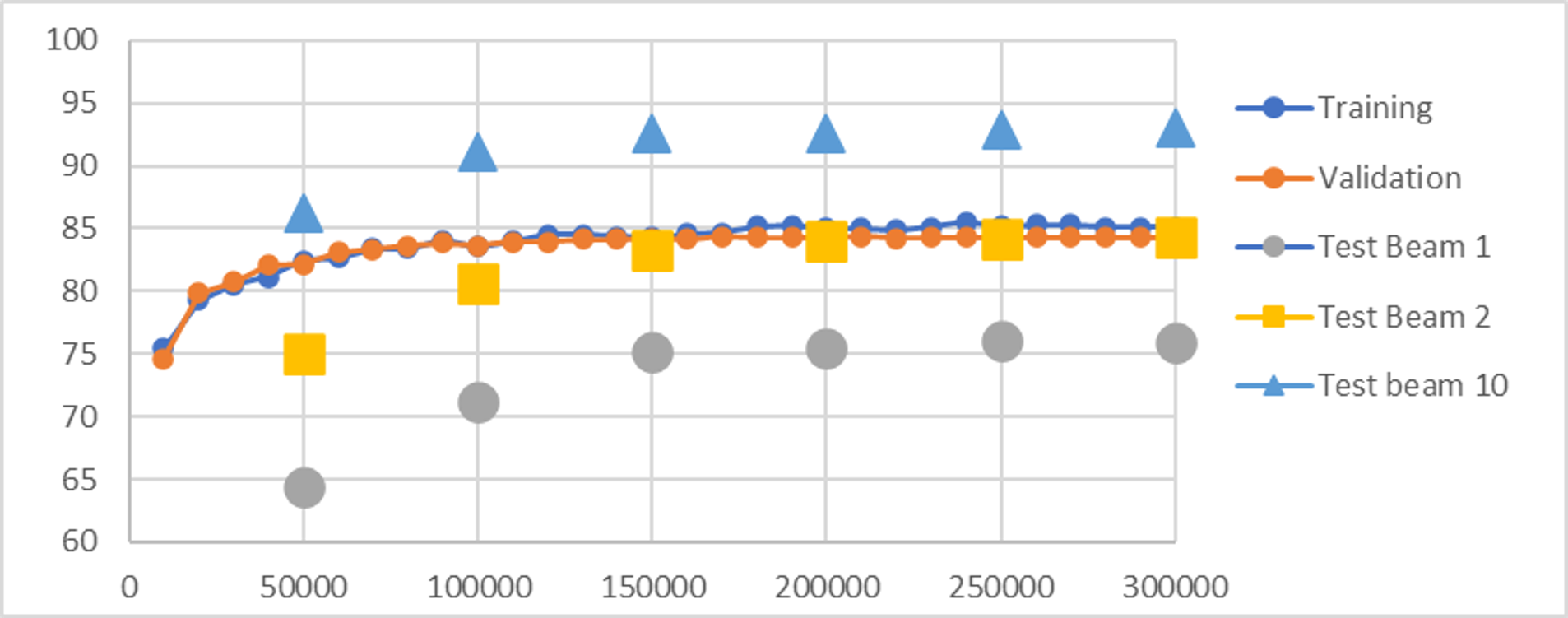}
\caption{Model training percentage accuracy up to 300,000 iterations on AxiomStep10. Training and Validation accuracies are per-token on the target axioms in the samples. Test accuracies are for full correct proofs of P1 to P2.}
\label{fig:training}
\end{figure}

\paragraph*{Training convergence}
Since our model trains on axiomatic proofs which may vary in order (allowing 2 or 3 options to be correct and occur in the training set), we see our training and token accuracies plateau below 90\% during training for AxiomStep10 as shown in Figure~\ref{fig:training}. Full testset proof accuracies for beam width 10 exceed 90\%, but also plateau along with the training and validation results. This result differs from our WholeProof10 training, which achieves training and validation accuracies above 96\% because the expected axiom sequence is more predictable, but as we have seen less generalized. 

As another observation on generalization and overfitting, we note that figure~\ref{fig:training} shows a slight separation between the training and validation accuracies starting at around iteration 180,000. While the training accuracy rises slowly, validation accuracy plateaus, indicating slight overfitting on the training data. Yet our model continues to slowly increase in quality, with the model snapshot that scores best on both validation and test accuracies occurring at iteration 300,000. This is our golden model, with 93.1\% of P1 to P2 proofs accurately found using beam width 10.

\paragraph*{Testing simpler models}
In addition to the sequence-to-sequence and graph-to-sequence models, we
explored a feed-forward equal/not equal classifier on a simple version of
our language. That model uses an autoencoder on the program to find an
embedding of the program and then a classifier based on the program embeddings
found. It achieves a 73\% accuracy on identifying equivalent pairs in the test data, which, as expected, is much lower than the full proof rate of 93\% achieved with a graph-to-sequence proof generator on our full language. This simple experiment highlights the importance of a system which prevents the false positives which a classifier might have by creating a verifiable proof.

We explore initial language generation using a simple language in order to assess feasibility of different approaches. For fine tuning network parameters and architectural features, we add more complexity to the language as shown in table \ref{tab:ResultLang2}. Language IDs 1 through 3 are all based on a simple grammar which only allows the "+" or "-" operators on scalar variables labeled a through j. The only axiom is \texttt{Commute}, which can be applied on up to 3 nodes in language IDs 2 and 3. Language ID 4 adds the scalar constants 0 and 1, scalar operations * and /, and 4 more axioms. We perform a fair amount of network development on this model in an effort to maintain high accuracy rates. Language ID also 4 expands the operands to 3 types and hence the number of operators also increases. To speed up model evaluation, we reduced the program length for IDs 5, 6, and 7, allowing us to train larger data sets for more epochs. ID 7 is a forward looking-model which makes a minor increment to the language to support the analysis of loop rolling and unrolling, discussed further in section~\ref{sec:appendix:backedge}. ID 8 is the WholeProof5 model in relation to these early experiments.

\begin{table*}[h]
  \scriptsize
{\begin{tabular}{ |p{0.4cm}|p{6cm}||p{0.4cm}|p{0.4cm}|p{0.4cm}|p{0.7cm}|p{0.7cm}||p{0.7cm}|p{1cm}||p{0.7cm}|p{0.7cm}| }
 \hline
 ID & Description & \rotatebox{90}{\# Operators} & \rotatebox{90}{\# Axioms} & \rotatebox{90}{\# Operands} & \rotatebox{90}{Program length} & \rotatebox{90}{Rewrite rules length} & \rotatebox{90}{\parbox{2.9cm}{Graph2seq (G2S) \\ or seq2seq (S2S)}} & \rotatebox{90}{Training set size} & \rotatebox{90}{\parbox{2.9cm}{Percent matching \\ with beam width 1}} & \rotatebox{90}{\parbox{2.9cm}{Percent matching \\ with beam width 10}} \\
 \hline
 1 & Rewrite sequence is only single Commute, uses sequence-to-sequence model & 2 & 1 & 10 & 3-19 & 1-5 & S2S & 80,000 & 90.0\% & 96.2\% \\
 \hline
 2 & Rewrite sequence is exactly 2 Commutes, uses sequence-to-sequence model & 2 & 1 & 10 & 5-24 & 3-10 & S2S & 80,000 & 80.3\% & 96.5\% \\
 \hline
 3 & Rewrite sequence exactly 2 Commutes & 2 & 1 & 10 & 5-24 & 3-10 & G2S & 80,000 & 98.9\% & 99.8\% \\
 \hline
 4 & Rewrite sequence exactly 3 Commutes & 2 & 1 & 10 & 7-45 & 5-15 & G2S & 80,000 & 91.4\% & 99.0\% \\
 \hline
 5 & Rewrite sequence 1 to 3 Commutes & 2 & 1 & 10 & 3-45 & 1-15 & G2S & 180,000 & 97.1\% & 99.2\% \\
 \hline
 7 & Commute, Noop, Cancel, Distribute Left, Distribute Right & 4 & 5 & 12 & 3-45 & 1-15 & G2S & 180,000 & 93.1\% & 97.4\% \\
 \hline
 8 & Scalars, Vectors, and Matrixes & 16 & 5 & 20 & 3-30 & 1-25 & G2S & 250,000 & 88.3\% & 95.6\% \\
 \hline
 9 & 13 Axioms & 16 & 13 & 20 & 3-30 & 1-25 & G2S & 400,000 & 85.5\% & 95.5\% \\
 \hline
 10 & Rewrite sequence or Not\_equal & 16 & 13 & 20 & 3-30 & 1-25 & G2S & 500,000 & 79.8\% & 93.8\% \\
 \hline
 11 & Test sequence-to-sequence & 16 & 13 & 20 & 3-30 & 1-25 & S2S & 400,000 & 59.8\% & 81.1\% \\
 \hline
 12 & Add loop axioms & 18 & 15 & 20 & 3-30 & 1-25 & G2S & 400,000 & 83.8\% & 94.7\% \\
 \hline
\end{tabular}}
\caption{Results for various language complexities studied, on non-incremental models (WholeProof).}
\label{tab:ResultLang2}
\vspace{-.8cm}
\end{table*}

We designed our datasets in section~\ref{sec:samplegen} with the goal of using the varied models to understand the generalizability of \texttt{pe-graph2axiom} and to show that our model is not overfitting on training data. For these next experiments, all results of for beam width 10, which provides for a neural-network directed search of up to 10 axiomatic proofs of equivalence for each program pair. Recall that our most complex dataset is AxiomStep10 which includes $(P1,P2,S)$ samples requiring up to 10 rewrite rules, $P1$ and $P2$ can have up to 50 AST nodes each, and an AST depth of up to 7. AxiomStep5 has samples requiring up to 5 rewrite rules, $P1$ and $P2$ can have up to 25 AST nodes each, and an AST depth of up to 6. Tables~\ref{tab:newnodes} and~\ref{tab:appendix:newtreedepth} (repeated from main paper below) demonstrate the ability of a model trained on AxiomStep5 to perform well on the larger distribution of programs from AxiomStep10, implying that the model has generalized well to our program equivalence problem and that \texttt{pe-graph2axiom} does not overfit its response to merely the training set distribution.

\begin{table*}[h!tb]
  \caption{Generalizing to longer P1 inputs. Percentage pass rates for equivalence proofs with P1 having increasing program graph nodes. The model trained with the AxiomStep5 dataset had no training examples more than 25 program graph nodes yet it performs relatively well on these more complex problems. The furthest right column shows the \texttt{pe-graph2axiom} model results on the most complex dataset.}
  \label{tab:newnodes}
  \small
  \centering
  \begin{tabular}{@{}rrrrrrrrrr@{}}
    \toprule
    & & \multicolumn{2}{c}{Testset} & & \multicolumn{2}{c}{Model trained} & & \multicolumn{2}{c}{Model trained}  \\
    & & \multicolumn{2}{c}{Sample Count} & & \multicolumn{2}{c}{on AxiomStep5} & & \multicolumn{2}{c}{on AxiomStep10}  \\
    \cmidrule{3-4} \cmidrule{6-7} \cmidrule{9-10}
    P1 nodes & & AS5 & AS10 & &  AS5 & AS10 & & AS5 & AS10 \\
    \midrule
    1-5 & & 231 & 109 & & 100 & 100 & & 100 & \textbf{100} \\
    6-10 & & 2147 & 1050 & & 100 & 99 & & 99 & \textbf{99} \\
    11-15 & & 3980 & 2175 & & 99 & 96 & & 99 & \textbf{96} \\
    16-20 & & 2583 & 2327 & & 98 & 92 & & 98 & \textbf{93} \\
    21-25 & & 1059 & 1989 & & 97 & 89 & & 98 & \textbf{92} \\
    26-30 & & 0 & 1229 & & N/A & 83 & & N/A & \textbf{90} \\
    31-35 & & 0 & 698 & & N/A & 78 & & N/A & \textbf{88} \\
    36-40 & & 0 & 304 & & N/A & 74 & & N/A & \textbf{87} \\
    41-45 & & 0 & 101 & & N/A & 68 & & N/A & \textbf{84} \\
    46-50 & & 0 & 27 & & N/A & 67 & & N/A & \textbf{85} \\
    All & & 10000 & 10000 & & 99 & 90 & & 99 & \textbf{93} \\
    \bottomrule
  \end{tabular}
\end{table*}

\begin{table*}[h!tb]
\caption{\label{tab:appendix:newtreedepth}Performance vs. AST size: counts and percentage pass rates.}
%  \caption{Study on generalizing to larger AST sizes. When considering the 10,000 tests from each of the AxiomStep5 set (AS5) and AxiomStep10 (AS10), columns show sample counts and percent pass rates for P1 having increasing AST depths. The model trained with the AxiomStep5 dataset had no training examples with AST depths of 7 yet it scales well to these more complex problems.}
  \small
  \centering
%\vspace{-.5cm}
\begin{tabular}{@{}rrrrrrrrrr@{}}
    \toprule
    & & \multicolumn{2}{c}{Testset} & & \multicolumn{2}{c}{Model trained} & & \multicolumn{2}{c}{Model trained}  \\
    & & \multicolumn{2}{c}{Sample Count} & & \multicolumn{2}{c}{on AxiomStep5} & & \multicolumn{2}{c}{on AxiomStep10}  \\
    \cmidrule{3-4} \cmidrule{6-7} \cmidrule{9-10}
    AST depth & & AS5 & AS10 & &  AS5 & AS10 & & AS5 & AS10 \\
    \midrule
    2 & & 5 & 3 & & 100 & 100 & & 100 & 100 \\
    3 & & 306 & 133 & & 100 & 100 & & 100 & 100 \\
    4 & & 1489 & 577 & & 100 & 99 & & 99 & 99 \\
    5 & & 4744 & 1844 & & 99 & 94 & & 98 & 95 \\
    6 & & 3456 & 4308 & & 98 & 90 & & 98 & 93 \\
    7 & & 0 & 3135 & & n/a & 86 & & n/a & 92 \\
    All & & 10000 & 10000 & & 99 & 90 & & 99 & \textbf{93} \\
    \bottomrule
  \end{tabular}
%  \vspace{-.3cm}
\end{table*}

Table~\ref{tab:appendix:newtreedepth} illustrates the ability of a model trained on AxiomStep5 (i.e., limited to proofs of length 5) to perform well when evaluated on the more complex AxiomStep10, which includes proofs of unseen length of up to 10. The robustness to the input program complexity is illustrated with the 86\% pass rate on AST depth 7, for the model trained on AxiomStep5 which never saw programs of depth 7 during training.

As an indication of the breadth of equivalent programs represented by AxiomStep10 relative to WholeProof10, table \ref{tab:modeltest:appendix} shows the full detail of models trained on all 4 datasets when tested on test data from all 4 datasets. AxiomStep10, while training on our broadest dataset in which axioms can be applied to nodes repeatedly and in variable order, achieves a 93\% average success rate. 72\% of the proofs of length 6 from the WholeProof10 testset were solved by the model trained on WholeProof10, but only 5\% of such proofs from AxiomStep10 were, suggesting the method of generating AxiomStep pairs covers the problem space more thoroughly. 

The complete result for the WholeProof10 model on the WholeProof10 dataset was 8,388 out of 10,000 program pairs had a correct proof found; of those, 8,350 were the exact proof created during $P1,P2$ generation, implying that WholeProof10, while performing well on its own testset distribution, is not learning to generalize to alternative proof paths. 

\begin{table*}[h!tb]
  \caption{Generalizing to longer proofs. Percentage pass rates for equivalence proofs of increasing axiom counts when testing each of 4 datasets on models trained using each of 4 datasets.}
  \label{tab:modeltest:appendix}
  \small
  \centering
  \setlength\tabcolsep{2pt}
  \begin{tabular}{@{}r|rrrrrrrrrrrrrrrrrrrr@{}}
    \toprule
    Axiom & & \multicolumn{4}{c}{Model trained on} & & \multicolumn{4}{c}{Model trained on} & & \multicolumn{4}{c}{Model trained on} & & \multicolumn{4}{c}{Model trained on}   \\
    Count in & & \multicolumn{4}{c}{WholeProof5 (WP5)} & & \multicolumn{4}{c}{WholeProof10 (WP10)} & & \multicolumn{4}{c}{AxiomStep5 (AS5)} & & \multicolumn{4}{c}{AxiomStep10 (AS10)}   \\
    \cmidrule{3-6} \cmidrule{8-11} \cmidrule{13-16} \cmidrule{18-21}
    Proof & & WP5 & \scriptsize{WP10} & AS5 & \scriptsize{AS10} & & WP5 & \scriptsize{WP10} & AS5 & \scriptsize{AS10} & & WP5 & \scriptsize{WP10} & AS5 & \scriptsize{AS10} & & WP5 & \scriptsize{WP10} & AS5 & \scriptsize{AS10} \\
    \midrule
    1 & & 100 & 100 & 100 & 99 & & 100 & 100 & 100 & 100 & & 100 & 100 & 100 & 100 & & 100 & 100 & 100 & \textbf{100} \\
    2 & & 99 & 98 & 66 & 64 & & 99 & 99 & 65 & 63 & & 100 & 99 & 100 & 99 & & 100 & 100 & 100 & \textbf{100} \\
    3 & & 98 & 94 & 34 & 33 & & 97 & 95 & 33 & 33 & & 100 & 98 & 99 & 98 & & 100 & 99 & 99 & \textbf{99} \\
    4 & & 93 & 84 & 16 & 15 & & 90 & 88 & 16 & 15 & & 98 & 95 & 98 & 97 & & 99 & 98 & 98 & \textbf{98} \\
    5 & & 84 & 70 & 8 & 7 & & 84 & 82 & 8 & 7 & & 96 & 91 & 96 & 95 & & 97 & 95 & 96 & \textbf{96} \\
    \midrule
    6 & &   & 14 &   & 4 & &   & 72 &   & 5 & &   & 81 &   & 88 & &   & 90 &   & \textbf{93} \\
    7 & &   & 0 &   & 1 & &   & 63 &   & 2 & &   & 67 &   & 81 & &   & 83 &   & \textbf{87} \\
    8 & &   & 0 &   & 0 & &   & 54 &   & 1 & &   & 54 &   & 75 & &   & 73 &   & \textbf{82} \\
    9 & &   & 0 &   & 0 & &   & 47 &   & 0 & &   & 35 &   & 64 & &   & 63 &   & \textbf{74} \\
    10 & &   & 0 &   & 0 & &   & 34 &   & 0 & &   & 24 &   & 57 & &   & 46 &   & \textbf{66} \\
    All & & 95 & 66 & 44 & 27 & & 94 & 84 & 44 & 27 & & 99 & 87 & 99 & 90 & & 99 & 93 & 99 & \textbf{93} \\
    \bottomrule
  \end{tabular}
\end{table*}

\paragraph*{Manual verifications} We conducted a series of manual verifications of the system used to produce all the above results. First, we are happy to confirm that most likely $AB \ne BA$ given no verifiable equivalence sequence was produced, but that provably $ab=ba$ indeed. We also verified that $A^{t^t}(B+C-C) = AB$, and that $AB\vec v -AB\vec w=AB(\vec v-\vec w)$ which would be a much faster implementation. The system correctly suggests that $AB\vec v -BA\vec w\ne AB(\vec v-\vec w)$. We ensured that $A^t(AA^t)^{-1}A\ne A^t(AA^{-1})^tA$, from a typo we once made when typing the computation of an orthonormal sub-space. We also verified that indeed $AB + AC + aD - aD = A(B+C)$.

\paragraph*{Generalizing variable types}
We explored the ability of the model to understand variable typing by training a model with the AxiomStep10 distribution but with no samples that included the scalar variable 'e' and scalar multiplication $*_s$. This removed about 50\% of the training set, as longer programs were often included both tokens. When tested with the unaltered AxiomStep10 test set and beam width 10, test samples that included a scalar variable not 'e' and $*_s$ were proven equal 90\% of the time; test samples that included 'e' and $*_s$ were also proven equal 90\% of the time. For beam width 1 the proof success rates were 72\% and 70\% for without and with 'e', implying that the heavily biased training set did have a small effect on the system generalization. \texttt{pe-graph2axiom} was still able to generalize the relation of 'e' to the $*_s$ operator given that 'e' was used in contexts similar to other scalar variables in the training samples that were provided, implying it was forming an internal representation of a 'scalar' type by learning from examples.

\subsection{Learning that multiple axiom choices are possible}
\label{sec:appendix:multiaxiom}

Our AxiomStep10 model is trained on axioms which may be applied in varying order in the training set. For example, $((a+b)*(c+d))=((b+a)*(d+c))$ may have the training data to Commute the left node $a+b$ first and then $c+d$ second; in the same dataset, $((a+e)*(b+c))=((e+a)*(c+b))$ might occur and the training data has the right node Commuted first. In this way, we expect the model to learn that either commuting the left or right node is a proper first axiom choice. Table~\ref{tab:beam3} explores the ability of the model to produce such axiom proposals. Given 5 scalar variables, there are 120 possible expressions where two 2-variable additions are multiplied together such as $((a+b)*(c+d))$. We consider here all 120 program pairs in which the left and right additions are commuted. The table shows which axioms and positions are recommended by the graph-to-sequence neural network model within the \texttt{pe-graph2axiom} system as most probably moving the 2 programs closer to equivalence by the beam width 3 on this problem. Note that the 2 correct axioms are always within the top 3 choices and the other 2 axioms (Commute and DistributeLeft on the root), while not necessary for this problem, are at least legal choices for axioms within our expression language.

The results in table~\ref{tab:beam3} relate to the value of our approach in relation to reinforcement learning models for proof generation  \cite{Alhussein19} \cite{Bansal19}. To make an analogy with reinforcement learning, in our training, the world 'state' is presented as a $P1,P2$ pair and the system must learn to produce an axiom at a location which performs an 'action' on the 'state' of $P1$ in a predictable way. Unlike reinforcement learning, we do not produce a reward function and our system cannot learn from a poor reward produced by an incorrect axiom. However, we have demonstrated that our system, as it is presented with a wide distribution of $(P1,P2,S)$ tuples to train on, learns a probability distribution of possibly correct axioms to produce for a given program pair. There may be value in combining our graph-neural-network within a reinforcement learning framework that used a hindsight mechanism \cite{Andrychowicz17} to learn from every attempted axiom, but it is not immediately obvious that our approach of learning only from examples of successful equivalence proofs would be improved.

\begin{table}[h!tb]
  \caption{Learning multiple output options. When considering scalar expressions that can be proven equivalent by commuting the left and right subexpressions, such as $(a+b)(c+d)=(b+a)(d+c)$, \texttt{pe-graph2axiom} learns that either the left or right commute can occur first. The columns show counts for axioms and locations proposed by the token generator with beam width of 3 when given 120 different scalar expression pairs.}
  \label{tab:beam3}
  \small
  \centering
  \begin{tabular}{@{}lrrrrr@{}}
    \toprule
    & & \multicolumn{4}{c}{Axiom} \\
    Beam & & Commute & Commute & Commute & DistributeLeft \\
    position & & left child & right child & root & root   \\
    \cmidrule{1-1} \cmidrule{3-6}
    First & & 49 & 35 & 36 & 0 \\
    Second & & 58 & 59 & 3 & 0 \\
    Third & & 13 & 26 & 45 & 36 \\
    Any of top 3 & & 120 & 120 & 84 & 36 \\
    \bottomrule
  \end{tabular}
\end{table}

\paragraph*{Exploration of alternate designs}
In order to design the system, we explored parts of the design space quickly and performed several single training run comparisons between 2 options, as shown in Table~\ref{tab:ParamSearch}. 

In cases where 2 options were similar, we
chose the model which ran faster, or run the models a second time to
get a more precise evaluation, or use our experience based on prior
experiments to select an option.

\begin{table}[h!tb]
\caption{Example explorations as a single feature or parameter is changed. Each comparison is a distinct experiment, as the entire network and language used was being varied.}
\label{tab:ParamSearch}
\centering\small
{\begin{tabular}{ |p{5.2cm}|p{1cm}|p{1.2cm}| }
 \hline
                  & Match  & Match   \\
 Options compared & beam 1 & beam 10 \\
 \hline
 \hline
 1 layer LSTM vs  &  198 & 1380 \\
 2 layer LSTM vs  & 5020 & 9457 \\
 3 layer LSTM     & 4358 & 8728 \\
 \hline
 No edges to grandchild nodes vs      & 9244 & 9728 \\
 Edges to grandchild nodes            & 9284 & 9774 \\
 \hline
 Encoder->Decoder only root node vs      & 8616 & 9472 \\
 Encoder->Decoder avg all nodes          & 7828 & 9292 \\
 \hline
\end{tabular}}
%\vspace{-.8cm}
\end{table}

Experiments such as these informed our final network architecture. For 
example, in \texttt{pe-graph2axiom}, we include 4 edges with learnable weight
matrices from a node to its grandchildren because such edges were found to
improve results on multiple runs.
Li et al. \cite{Li19}  discusses the importance of selecting the optimal process for aggregating
the graph information hence we explore that issue for our network.
Our approach uses the root comparison
node to create aggregate the graph information for the decoder as it performs
better than a node averaging. 

\paragraph*{Including Not\_equal option}

Table \ref{tab:Proof} analyzes the challenge related to a model which only predicts
Equal or Not\_equal for program pairs along with various options which produce
rewrite rules which can be checked for correctness. In all 4 output cases shown,
2 programs are provided as input. These programs use an earlier version of our language model with 16 operators, 13 core axioms, and 20 operands generated with a distribution similar to WholeProof5.

\begin{table}[h!tb]
\caption{Table showing alternate options for handling not equal programs\label{tab:Proof}}
\centering{\small
\begin{tabular}{ |p{2.2cm}|p{0.8cm}||p{1.2cm}|p{1.2cm}|p{1cm}| }
 \hline
 Network     &        &           & Predicted & Correct \\
 output      &        & Predicted & Rules     & Rewrite \\
 Description & Actual & NotEq     & or Eq     & Rules   \\
 \hline
 \hline
 Eq or NotEq,    & Eq    & 5.4\%  & 94.6\% & N/A \\
 Beam width 1    & NotEq & 90.4\% & 9.6\% & N/A \\
 \hline
 Rules or NotEq, & Eq    & 6.6\% & 93.4\% & 70.7\% \\
 Beam width 1    & NotEq & 90.9\% & 9.1\% & N/A \\
 \hline
 Rules only,     & Eq    & N/A & 100\% & 87.8\% \\
 Beam width 1    & NotEq & N/A & N/A & N/A \\
 \hline
 Rules only,     & Eq    & N/A & 100\% & 96.2\% \\
 Beam width 10   & NotEq & N/A & N/A & N/A \\
 \hline
\end{tabular}
}
\end{table}

For the first output case, the output
sequence to produce is either \texttt{Equal} or \texttt{Not\_equal}. Given a
false positive rate of 9.6\%, these results
demonstrate the importance of producing a verifiable proof of equivalence
when using machine learning for automated equivalence checking.
For the second output case, the model can produce either \texttt{Not\_equal}
or a rewrite rule sequence which can be checked for correctness. The source
programs for the first and second case are identical: 250,000 equivalent program pairs
and 250,000 non-equivalent program pairs. In the second case, the false positive
rate from the network is 9.1\% (rules predicted for Not\_equal programs), but
the model only produces correct rewrite rules between actual equivalent programs
in 70.7\% of the cases. 

One challenge with a model that produce rules or
\texttt{Not\_equal} is that beam widths beyond 1 are less usable. Consider that
with a beam width of 1, if the network predicts \texttt{Not\_equal} then the
checker would conclude the programs are not equal (which is
correct for 90.9\% of the actually not equal programs). With a beam width of 10,
there would be more proposed rewrite rules for equal programs to test with, but
if 1 of the 10 proposals is \texttt{Not\_equal}, should the checker conclude they
are not equal? Or should the the checker only consider the most likely prediction
(beam width 1) when checking for non-equivalence? The third and fourth network 
output cases provide an answer. For these 2 cases, the training set is 400,000 
equivalent program pairs - none are non-equivalent. 250,000 of these pairs are
identical to the equivalent programs in the first 2 cases, and 150,000 are new
but were produced using the same random generation process. Note that by requiring
the network to focus only on creating rewrite rules, beam width 1 is able to 
create correct rewrite rules for 87.8\% of the equivalent programs. And now,
since we've remove the confusion of the \texttt{Not\_equal} prediction option,
beam width 10 can be used to produce 10 possible rewrite rule sequences and
in 96.2\% of the cases these rules are correct. Hence, we propose the preferred
use model for pe-graph2axiom is to always use the model which is trained for
rule generation with beam width 10 and rely on our rule checker to prevent
false positives. From the 10 rewrite rule proposals, non-equivalent programs
will never have a correct rewrite rule sequence produced, hence we
guarantee there are no false positives.

%%% LNP: I propose to comment out this entire subsec. Remove the \begin/end{comment} to uncomment
\subsection{An Example of Back-Edge in the Program Graph}
\label{sec:appendix:backedge}

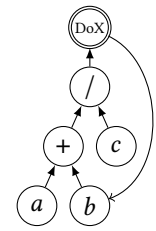
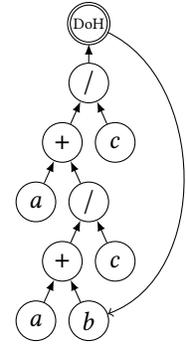
\begin{figure}[h!tb]
  %% \vspace{-.3cm}
\centering
    \begin{minipage}[b]{0.25 \textwidth}
        \centering
        \subfloat [DoX($b=(a+b)/c$)] {
            \begin{tikzpicture} [level distance=2.25em, inner sep=1pt, minimum size=1.5em, sibling distance=2em, edge from parent/.style={draw,latex-}]
                \node [circle, double, draw] (root) {\tiny DoX}
                    child {node [circle, draw] {$/$}
                        child{node [circle, draw] {$+$}
                            child {node [circle, draw] {$a$}}
                            child {node [circle, draw] (leaf) {$b$}}
                        }
                        child {node [circle, draw] {$c$}
                            child{edge from parent[draw=none] node [opacity=0] {}}
                            child{edge from parent[draw=none] node [opacity=0] {}}
                        }
                    };
                \path [->, draw] (root) to [out=335, in=20] (leaf);
            \end{tikzpicture}
        }
    \end{minipage}
    \hfill
    \unskip\ \vrule\
    \begin{minipage}[b]{0.2 \textwidth}
        \centering
        \subfloat [DoHalf($b=(a+(a+b)/c)/c$)] {
            \begin{tikzpicture} [level distance=2.25em, inner sep=1pt, minimum size=1.5em, sibling distance=2em, edge from parent/.style={draw,latex-}]
                \node [circle, double, draw] (root) {\tiny DoH}
                    child {node [circle, draw] {$/$}
                        child{node [circle, draw] {$+$}
                            child {node [circle, draw] {$a$}}
                            child {node [circle, draw] {$/$}
                                child{node [circle, draw] {$+$}
                                    child {node [circle, draw] {$a$}}
                                    child {node [circle, draw] (leaf) {$b$}}
                                }
                                child {node [circle, draw] {$c$}}
                            }
                        }
                        child {node [circle, draw] {$c$}}
                    };
                \path [->, draw] (root) to [out=335, in=20] (leaf);
            \end{tikzpicture}
        }
    \end{minipage}
\caption{Adding loop constructs creates cycles in the program graph.}
\label{fig:LoopAST}
\end{figure}

Figure~\ref{fig:LoopAST} shows an example of DoX and DoHalf.
The new operators result in 2 new edges in our graph representation (along with 2 new back-edges): there is a 'loopbody' edge type from the loop operator node to the start of the subgraph, and there is a 'loopfeedback' edge type from the variable which is written to each loop iteration. These 2 edge types are shown in the figure. The new $Dohalf$ axiom intuitively states that $DoX(g(y)) = DoHalf(g(g(y)))$ (where $y$ is the variable reused each iteration), and $Dox$ states the reverse. 
%% \FloatBarrier

%% {\bf The full axiom list is presented in the next pages}

%% \clearpage
%% \FloatBarrier

%% Text of appendix \ldots

\end{document}